\input harvmac.tex
\input epsf.tex
\input amssym
\input ulem.sty


\let\includefigures=\iftrue
\let\useblackboard=\iftrue
\newfam\black

\def\figin{\epsfcheck\figin}\def\figins{\epsfcheck\figins}
\def\epsfcheck{\ifx\epsfbox\UnDeFiNeD
\message{(NO epsf.tex, FIGURES WILL BE IGNORED)}
\gdef\figin##1{\vskip2in}\gdef\figins##1{\hskip.5in}
\else\message{(FIGURES WILL BE INCLUDED)}%
\gdef\figin##1{##1}\gdef\figins##1{##1}\fi}
\def\DefWarn#1{}
\def\figinsert{\goodbreak\midinsert}
\def\ifig#1#2#3{\DefWarn#1\xdef#1{fig.~\the\figno}
\writedef{#1\leftbracket fig.\noexpand~\the\figno} %
\figinsert\figin{\centerline{#3}}\medskip\centerline{\vbox{\baselineskip12pt
\advance\hsize by -1truein\noindent\footnotefont{\bf
Fig.~\the\figno:} #2}}
\bigskip\endinsert\global\advance\figno by1}


\includefigures
\message{If you do not have epsf.tex (to include figures),}
\message{change the option at the top of the tex file.}
\input epsf
\def\figin{\epsfcheck\figin}\def\figins{\epsfcheck\figins}
\def\epsfcheck{\ifx\epsfbox\UnDeFiNeD
\message{(NO epsf.tex, FIGURES WILL BE IGNORED)}
\gdef\figin##1{\vskip2in}\gdef\figins##1{\hskip.5in}
\else\message{(FIGURES WILL BE INCLUDED)}%
\gdef\figin##1{##1}\gdef\figins##1{##1}\fi}
\def\DefWarn#1{}
\def\figinsert{\goodbreak\midinsert}
\def\ifig#1#2#3{\DefWarn#1\xdef#1{fig.~\the\figno}
\writedef{#1\leftbracket fig.\noexpand~\the\figno}%
\figinsert\figin{\centerline{#3}}\medskip\centerline{\vbox{
\baselineskip12pt\advance\hsize by -1truein
\noindent\footnotefont{\bf Fig.~\the\figno:} #2}}
\endinsert\global\advance\figno by1}
\else
\def\ifig#1#2#3{\xdef#1{fig.~\the\figno}
\writedef{#1\leftbracket fig.\noexpand~\the\figno}%
\global\advance\figno by1} \fi

\def\figin{\epsfcheck\figin}\def\figins{\epsfcheck\figins}
\def\epsfcheck{\ifx\epsfbox\UnDeFiNeD
\message{(NO epsf.tex, FIGURES WILL BE IGNORED)}
\gdef\figin##1{\vskip2in}\gdef\figins##1{\hskip.5in}
\else\message{(FIGURES WILL BE INCLUDED)}%
\gdef\figin##1{##1}\gdef\figins##1{##1}\fi}
\def\DefWarn#1{}
\def\figinsert{\goodbreak\midinsert}
\def\ifig#1#2#3{\DefWarn#1\xdef#1{fig.~\the\figno}
\writedef{#1\leftbracket fig.\noexpand~\the\figno} %
\figinsert\figin{\centerline{#3}}\medskip\centerline{\vbox{\baselineskip12pt
\advance\hsize by -1truein\noindent\footnotefont{\bf
Fig.~\the\figno:} #2}}
\bigskip\endinsert\global\advance\figno by1}

\def \pa {\partial}

\def\OO{{\cal OO}}

\catcode`\@=11
\def\slash#1{\mathord{\mathpalette\c@ncel{#1}}}
\overfullrule=0pt

\def\CC{{\cal C}}
\def\DD{{\cal D}}

\def\FF{{\cal F}}
\def\GG{{\cal G}}

\def\NN{{\cal N}}
\def\OO{{\cal O}}

\def\WW{{\cal W}}

\def\underrel#1\over#2{\mathrel{\mathop{\kern\z@#1}\limits_{#2}}}

\catcode`\@=12

\def\lzone{{\underrel{\approx}\over{z\rightarrow 1}}}


\def\exp{{\rm exp}}


\def\zbar{{\bar z}}

\def\DD{{\cal D}}

\def\DL{{\Delta_{L}}}
\def\DH{{\Delta_{H}}}
\def\OL{{\OO_{L}}}
\def\OH{{\OO_{H}}}


\def\unlockat{\catcode`\@=11}
\def\lockat{\catcode`\@=12}

\unlockat

\def\newsec#1{\global\advance\secno by1\message{(\the\secno. #1)}
\global\subsecno=0\global\subsubsecno=0\eqnres@t\noindent
{\bf\the\secno. #1}
\writetoca{{\secsym} {#1}}\par\nobreak\medskip\nobreak}
\global\newcount\subsecno \global\subsecno=0
\def\subsec#1{\global\advance\subsecno
by1\message{(\secsym\the\subsecno. #1)}
\ifnum\lastpenalty>9000\else\bigbreak\fi\global\subsubsecno=0
\noindent{\it\secsym\the\subsecno. #1}
\writetoca{\string\quad {\secsym\the\subsecno.} {#1}}
\par\nobreak\medskip\nobreak}
\global\newcount\subsubsecno \global\subsubsecno=0
\def\subsubsec#1{\global\advance\subsubsecno by1
\message{(\secsym\the\subsecno.\the\subsubsecno. #1)}
\ifnum\lastpenalty>9000\else\bigbreak\fi
\noindent\quad{\secsym\the\subsecno.\the\subsubsecno.}{#1}
\writetoca{\string\qquad{\secsym\the\subsecno.\the\subsubsecno.}{#1}}
\par\nobreak\medskip\nobreak}

\def\subsubseclab#1{\DefWarn#1\xdef
#1{\noexpand\hyperref{}{subsubsection}%
{\secsym\the\subsecno.\the\subsubsecno}%
{\secsym\the\subsecno.\the\subsubsecno}}%
\writedef{#1\leftbracket#1}\wrlabeL{#1=#1}}
\lockat

\lref\OsbornCR{
  H.~Osborn and A.~C.~Petkou,
  ``Implications of conformal invariance in field theories for general dimensions,''
Annals Phys.\  {\bf 231}, 311 (1994).
[hep-th/9307010].
}
\lref\HuangHYE{
  K.~W.~Huang,
  ``$d>2$ Stress-Tensor OPE near a Line,''
[arXiv:2103.09930 [hep-th]].
}
\lref\AldayQKM{
  L.~F.~Alday, J.~B.~Bae, N.~Benjamin and C.~Jorge-Diaz,
  ``On the Spectrum of Pure Higher Spin Gravity,''
JHEP {\bf 2012}, 001 (2020).
[arXiv:2009.01830 [hep-th]].
}
\lref\GaberdielJCA{
  M.~R.~Gaberdiel, K.~Jin and E.~Perlmutter,
  ``Probing higher spin black holes from CFT,''
JHEP {\bf 1310}, 045 (2013).
[arXiv:1307.2221 [hep-th]].
}
\lref\AfkhamiJeddiIDC{
  N.~Afkhami-Jeddi, K.~Colville, T.~Hartman, A.~Maloney and E.~Perlmutter,
  ``Constraints on higher spin CFT$_{2}$,''
JHEP {\bf 1805}, 092 (2018).
[arXiv:1707.07717 [hep-th]].
}
\lref\WattsZQ{
  G.~M.~T.~Watts,
  ``Fusion in the W(3) algebra,''
Commun.\ Math.\ Phys.\  {\bf 171}, 87 (1995).
[hep-th/9403163].
}
\lref\PerlmutterPKF{
  E.~Perlmutter,
  ``Bounding the Space of Holographic CFTs with Chaos,''
JHEP {\bf 1610}, 069 (2016).
[arXiv:1602.08272 [hep-th]].
}

\lref\RasmussenEUS{
  J.~Rasmussen and C.~Raymond,
  ``Galilean contractions of $W$-algebras,''
Nucl.\ Phys.\ B {\bf 922}, 435 (2017).
[arXiv:1701.04437 [hep-th]].
}
\lref\HegdeDQH{
  A.~Hegde, P.~Kraus and E.~Perlmutter,
  ``General Results for Higher Spin Wilson Lines and Entanglement in Vasiliev Theory,''
JHEP {\bf 1601}, 176 (2016).
[arXiv:1511.05555 [hep-th]].
}
\lref\CastroIW{
  A.~Castro, R.~Gopakumar, M.~Gutperle and J.~Raeymaekers,
  ``Conical Defects in Higher Spin Theories,''
JHEP {\bf 1202}, 096 (2012).
[arXiv:1111.3381 [hep-th]].
}
\lref\deBoerSNA{
  J.~de Boer, A.~Castro, E.~Hijano, J.~I.~Jottar and P.~Kraus,
  ``Higher spin entanglement and $\WW_N$ conformal blocks,''
JHEP {\bf 1507}, 168 (2015).
[arXiv:1412.7520 [hep-th]].
}
\lref\PopeRev{
  C.~Pope,
  ``Lectures on $W$ algebras and $W$ gravity,''
Trieste HEP Cosmol.1991:827-867 .
[arXiv:9112076 [hep-th]].
}
\lref\ZhuMc{
  C.~J.~Zhu,
  ``The Complete structure of the nonlinear W(4) and W(5),''
Phys. Lett. B {\bf 316} (1993), 264-274.
[arXiv:9306025 [hep-th]].
}
\lref\FitzpatrickFOA{
  A.~L.~Fitzpatrick, J.~Kaplan, M.~T.~Walters and J.~Wang,
  ``Hawking from Catalan,''
JHEP {\bf 1605}, 069 (2016).
[arXiv:1510.00014 [hep-th]].
}

\lref\KulaxiziTKD{
  M.~Kulaxizi, G.~S.~Ng and A.~Parnachev,
  ``Subleading Eikonal, AdS/CFT and Double Stress Tensors,''
  JHEP {\bf 1910}, 107 (2019).
[arXiv:1907.00867 [hep-th]].
}
\lref\KarlssonDBD{
  R.~Karlsson, M.~Kulaxizi, A.~Parnachev and P.~Tadi\' c,
  ``Leading Multi-Stress Tensors and Conformal Bootstrap,''
JHEP {\bf 2001}, 076 (2020).
[arXiv:1909.05775 [hep-th]].
}
\lref\KomargodskiEK{
  Z.~Komargodski and A.~Zhiboedov,
  ``Convexity and Liberation at Large Spin,''
JHEP {\bf 1311}, 140 (2013).
[arXiv:1212.4103 [hep-th]].
}
\lref\FitzpatrickYX{
  A.~L.~Fitzpatrick, J.~Kaplan, D.~Poland and D.~Simmons-Duffin,
  ``The Analytic Bootstrap and AdS Superhorizon Locality,''
JHEP {\bf 1312}, 004 (2013).
[arXiv:1212.3616 [hep-th]].
}
\lref\FitzpatrickVUA{
  A.~L.~Fitzpatrick, J.~Kaplan and M.~T.~Walters,
  ``Universality of Long-Distance AdS Physics from the CFT Bootstrap,''
JHEP {\bf 1408}, 145 (2014).
[arXiv:1403.6829 [hep-th]].
}
\lref\FitzpatrickZHA{
  A.~L.~Fitzpatrick, J.~Kaplan and M.~T.~Walters,
  ``Virasoro Conformal Blocks and Thermality from Classical Background Fields,''
JHEP {\bf 1511}, 200 (2015).
[arXiv:1501.05315 [hep-th]].
}
\lref\FitzpatrickIVE{
  A.~L.~Fitzpatrick, J.~Kaplan, D.~Li and J.~Wang,
  ``On information loss in AdS$_{3}$/CFT$_{2}$,''
JHEP {\bf 1605}, 109 (2016).
[arXiv:1603.08925 [hep-th]].
}
\lref\AsplundRE{
  C.~T.~Asplund, A.~Bernamonti, F.~Galli and T.~Hartman,
  ``Holographic Entanglement Entropy from 2d CFT: Heavy States and Local Quenches,''
  JHEP {\bf 1502}, 171 (2015).
  [arXiv:1410.1392 [hep-th]].
}
\lref\KarlssonQFI{
  R.~Karlsson, M.~Kulaxizi, A.~Parnachev and P.~Tadi\' c,
  ``Black Holes and Conformal Regge Bootstrap,''
JHEP {\bf 1910}, 046 (2019).
[arXiv:1904.00060 [hep-th]].
}
\lref\LiTPF{
  Y.~Z.~Li, Z.~F.~Mai and H.~Lü,
  ``Holographic OPE Coefficients from AdS Black Holes with Matters,''
JHEP {\bf 1909}, 001 (2019).
[arXiv:1905.09302 [hep-th]].
}
\lref\FitzpatrickZQZ{
  A.~L.~Fitzpatrick and K.~W.~Huang,
  ``Universal Lowest-Twist in CFTs from Holography,''
  JHEP {\bf 1908}, 138 (2019).
[arXiv:1903.05306 [hep-th]].
}
\lref\KarlssonDBD{
  R.~Karlsson, M.~Kulaxizi, A.~Parnachev and P.~Tadi\' c,
  ``Leading Multi-Stress Tensors and Conformal Bootstrap,''
JHEP {\bf 2001}, 076 (2020).
[arXiv:1909.05775 [hep-th]].
}
\lref\LiZBA{
  Y.~Z.~Li,
  ``Heavy-light Bootstrap from Lorentzian Inversion Formula,''
JHEP {\bf 2007}, 046 (2020).
[arXiv:1910.06357 [hep-th]].
}
\lref\KarlssonGHX{
  R.~Karlsson, M.~Kulaxizi, A.~Parnachev and P.~Tadi\' c,
  ``Stress tensor sector of conformal correlators operators in the Regge limit,''
JHEP {\bf 2007}, 019 (2020).
[arXiv:2002.12254 [hep-th]].
}
\lref\LiDQM{
  Y.~Z.~Li and H.~Y.~Zhang,
  ``More on Heavy-Light Bootstrap up to Double-Stress-Tensor,''
  JHEP {\bf 2010}, 055 (2020).
[arXiv:2004.04758 [hep-th]].
}
\lref\FitzpatrickYJB{
  A.~L.~Fitzpatrick, K.~W.~Huang, D.~Meltzer, E.~Perlmutter and D.~Simmons-Duffin,
  ``Model-Dependence of Minimal-Twist OPEs in $d>2$ Holographic CFTs,''
  JHEP {\bf 2011}, 060 (2020).
[arXiv:2007.07382 [hep-th]].
}
\lref\KulaxiziDXO{
  M.~Kulaxizi, G.~S.~Ng and A.~Parnachev,
  ``Black Holes, Heavy States, Phase Shift and Anomalous Dimensions,''
SciPost Phys.\  {\bf 6}, 065 (2019).
[arXiv:1812.03120 [hep-th]].
}
\lref\KarlssonTXU{
  R.~Karlsson,
  ``Multi-stress tensors and next-to-leading singularities in the Regge limit,''
JHEP {\bf 2008}, 037 (2020).
[arXiv:1912.01577 [hep-th]].
}
\lref\ParnachevFNA{
  A.~Parnachev,
  ``Near Lightcone Thermal Conformal Correlators and Holography,''
J. Phys. A {\bf 54}, no.15, 155401 (2021).
[arXiv:2005.06877 [hep-th]].
}
\lref\ParnachevZBR{
 A.~Parnachev and K.~Sen,
  ``Notes on AdS-Schwarzschild eikonal phase,''
JHEP {\bf 2003}, 289 (2021).
[arXiv:2011.06920 [hep-th]].
}
\lref\FitzpatrickEFK{
  A.~L.~Fitzpatrick, K.~W.~Huang and D.~Li,
  ``Probing universalities in d > 2 CFTs: from black holes to shockwaves,''
JHEP {\bf 1911}, 139 (2019).
[arXiv:1907.10810 [hep-th]].
}
\lref\HuangFOG{
  K.~W.~Huang,
  ``Stress-tensor commutators in conformal field theories near the lightcone,''
Phys.\ Rev.\ D {\bf 100}, no. 6, 061701 (2019).
[arXiv:1907.00599 [hep-th]].
}
\lref\HuangHYE{
  K.~W.~Huang,
  ``$d>2$ stress-tensor operator product expansion near a line,''
Phys.\ Rev.\ D {\bf 103}, no. 12, 121702 (2021).
[arXiv:2103.09930 [hep-th]].
}
\lref\HijanoRLA{
  E.~Hijano, P.~Kraus and R.~Snively,
  ``Worldline approach to semi-classical conformal blocks,''
JHEP {\bf 1507}, 131 (2015).
[arXiv:1501.02260 [hep-th]].
}
\lref\HijanoQJA{
  E.~Hijano, P.~Kraus, E.~Perlmutter and R.~Snively,
  ``Semiclassical Virasoro blocks from AdS$_{3}$ gravity,''
JHEP {\bf 1512}, 077 (2015).
[arXiv:1508.04987 [hep-th]].
}

\lref\CotlerZFF{
  J.~Cotler and K.~Jensen,
  ``A theory of reparameterizations for AdS$_3$ gravity,''
JHEP {\bf 1902}, 079 (2019).
[arXiv:1808.03263 [hep-th]].
}
\lref\CollierEXN{
  S.~Collier, Y.~Gobeil, H.~Maxfield and E.~Perlmutter,
  ``Quantum Regge Trajectories and the Virasoro Analytic Bootstrap,''
JHEP {\bf 1905}, 212 (2019).
[arXiv:1811.05710 [hep-th]].
}
\lref\FitzpatrickDLT{
  A.~L.~Fitzpatrick and J.~Kaplan,
  ``Conformal Blocks Beyond the Semi-Classical Limit,''
JHEP {\bf 1605}, 075 (2016).
[arXiv:1512.03052 [hep-th]].
}
\lref\AnousBHC{
  T.~Anous, T.~Hartman, A.~Rovai and J.~Sonner, 
  ``Black Hole Collapse in the 1/c Expansion,''
JHEP  {\bf 1607}, 123 (2016)
[arXiv:1603.04856 [hep-th]].
}
\lref\FitzpatrickMJQ{
  A.~L.~Fitzpatrick and J.~Kaplan,
  ``On the Late-Time Behavior of Virasoro Blocks and a Classification of Semiclassical Saddles,''
JHEP {\bf 1704}, 072 (2017).
[arXiv:1609.07153 [hep-th]].
}
\lref\ChenNVBIP{
  H.~Chen, C.~Hussong, J.~Kaplan and D.~Li,
  ``A Numerical Approach to Virasoro Blocks and the Information Paradox,''
JHEP {\bf 1709}, 102 (2017).
[arXiv:1703.09727 [hep-th]].
}
\lref\FaulknerHLL{
  T.~Faulkner and H.~Wang,
  ``Probing beyond ETH at large $c$,''
JHEP {\bf 1806}, 123 (2018).
[arXiv:1712.03464 [hep-th]].
}
\lref\CaputaQE{
  P.~Caputa, J.~Simon, A.~Stikonas and T.~Takayanagi, 
  ``Quantum Entanglement of Localized Excited States at Finite Temperature,''
JHEP {\bf 1501}, 102 (2015).
[arXiv:1410.2287 [hep-th]].
}
\lref\ChenEE{
B.~Chen and J.~Wu,
``Holographic Entanglement Entropy For a Large Class of States in 2D CFT,''
JHEP {\bf 1609}, 015 (2016).
[arXiv:1605.06753 [hep-th]].
}
\lref\ChenESC{
B.~Chen, J.~Wu and J.~Zhang,
``Holographic Description of 2D Conformal Block in Semi-classical Limit,''
JHEP {\bf 1610}, 110 (2016).
[arXiv:1609.00801 [hep-th]].
}
\lref\HartmanEE{
  T.~Hartman, 
  ``Entanglement Entropy at Large Central Charge,''
[arXiv:1303.6955 [hep-th]].
}
\lref\FaulknerYIA{
  T.~Faulkner,
  ``The Entanglement Renyi Entropies of Disjoint Intervals in AdS/CFT,''
[arXiv:1303.7221 [hep-th]].
}

\lref\CardyIE{
  J.~L.~Cardy,
  ``Operator Content of Two-Dimensional Conformally Invariant Theories,''
Nucl. Phys. B {\bf 270}, 186-204 (1986).
}
\lref\HijanoFJA{
  E.~Hijano, P.~Kraus and E.~Perlmutter,
  ``Matching four-point functions in higher spin AdS$_3$/CFT$_2$,''
JHEP {\bf 1305}, 163 (2013).
[arXiv:1302.6113 [hep-th]].
}
\lref\CastroMZA{
  A.~Castro and E.~Llabr\'es,
  ``Unravelling Holographic Entanglement Entropy in Higher Spin Theories,''
JHEP {\bf 1503}, 124 (2015).
[arXiv:1410.2870 [hep-th]].
}
\lref\LISZEWSKA{
  E.~Liszewska and W.~Młotkowski,
  ``Some relatives of the Catalan sequence,''
Advances in Applied Mathematics {\bf 121}, (2020).
[arXiv:1907.10725 [math]].
}
\lref\KorchemskyLRCFT{
  G.~P.~Korchemsky and A.~Zhiboedov,
  ``On the light-ray algebra in conformal field theories,''
[arXiv:2109.13269 [hep-th]].
}
\lref\BelinLSR{
  A.~Belin, D.~M.~Hofman, G.~Mathys and M.~T.~Walters,
  ``On the stress tensor light-ray operator algebra,''
JHEP {\bf 2105}, 033 (2021).
[arXiv:2011.13862 [hep-th]].
}
\lref\BeskenSNX{
  M.~Be\c{s}ken, J.~De Boer and G.~Mathys,
  ``On Local and Integrated Stress-Tensor Commutators,''
JHEP {\bf 2021}, 148 (2020).
[arXiv:2012.15724 [hep-th]].
}
\lref\FitzpatrickQMA{
  A.~L.~Fitzpatrick, J.~Kaplan, M.~T.~Walters and J.~Wang,
  ``Eikonalization of Conformal Blocks,''
JHEP {\bf 1509}, 019 (2015).
[arXiv:1504.01737 [hep-th]].
}
\lref\BanderierRYT{
  C.~Banderier, P.~Marchal and M.~Wallner
  ``Rectangular Young tableaux with local decreases and the density method for uniform random generation,''
[arXiv:1805.09017 [cs]].
}
\lref\SloaneISS{
  N. J. A.~Sloane, 
  ``The On-Line Encyclopedia of Integer Sequences,''
published electronically at: http://oeis.org/

}

\lref\AlekseevShatashvili{
A.~Alekseev and S.~L.~Shatashvili,
``Path Integral Quantization of the Coadjoint Orbits of the Virasoro Group and 2D Gravity,''
Nucl. Phys. B {\bf 323} (1989), 719-733
}

\Title{
\vbox{\baselineskip8pt
}}
{\vbox{
\centerline{CFT correlators, $\WW$-algebras and }
\vskip.1in
\centerline{Generalized Catalan Numbers}
}}

\vskip.1in
 \centerline{
Robin Karlsson,${}^{a}$ Manuela Kulaxizi,${}^{a}$ Gim Seng Ng,${}^{a}$ Andrei Parnachev${}^{a}$ and Petar Tadi\' c${}^{b}$ \footnote{}{${}^{a}$ karlsson, manuela, parnachev $@$ maths.tcd.ie, nggimseng@gmail.com}   
\footnote{}{${}^{b}$ petar.tadic$@$yale.edu}} \vskip.1in
\centerline{${}^{a}$ \it School of Mathematics, Trinity College Dublin, Dublin 2, Ireland}


\centerline{${}^{b}$ \it Department of Physics, Yale University, New Haven, CT 06520, USA}

\vskip.4in \centerline{\bf Abstract}{ 
\noindent 
In two spacetime dimensions the Virasoro heavy-heavy-light-light (HHLL) vacuum block in a certain limit is governed by the Catalan numbers. The equation for their generating function can be generalized to a differential equation which the logarithm of the block satisfies. We show that a similar story holds for the HHLL $\WW_N$ vacuum blocks, where a suitable generalization of the Catalan numbers plays the main role. 
Moreover, the $\WW_N$ blocks have the same form as the stress tensor sector 
of HHLL near lightcone conformal correlators in $2(N-1)$ spacetime dimensions. In the latter case the Catalan numbers are generalized to the numbers of linear extensions of certain partially ordered sets. 
}
 
\Date{November 2021}

\listtoc\writetoc
\vskip 1.0in \noindent

\eject

\newsec{Introduction and summary of results}
\noindent The local conformal algebra in two dimensions is the infinite-dimensional Virasoro algebra and is generated by the modes of the stress tensor operator $T(z)$. It induces a natural decomposition of correlation functions into Virasoro conformal blocks which capture the contribution from a given Virasoro primary and all its Virasoro descendants. With respect to the global conformal algebra, each Virasoro representation contains an infinite number of quasi-primaries -- the Virasoro symmetry therefore imposes strong constraints on the theory as seen from the perspective of someone that only knew about its global part. 

Further, the presence of symmetries in CFTs is deeply connected to universal features. An example is Cardy's formula for the density of high energy of states in two-dimensional CFTs \CardyIE. It follows from the large conformal transformation of the torus and the dominance of the lowest dimension operator in the partition function in the low-temperature limit. Another example of universality is the presence of large-spin double-twist composite operators in any unitary $d>2$-dimensional CFT \refs{\KomargodskiEK,\FitzpatrickYX}. This follows from studying the lightcone limit of a four-point function of scalar operators and utilizing crossing symmetry. In one channel the identity operator dominates because it is the operator with the smallest twist $\tau=0$. Interpreting this in a different channel leads to the existence of double-twist operators with large spin $\ell$ and universal OPE data to leading order in the $\ell^{-1}$ expansion. Furthermore, in the absence of light scalars ($\Delta_{\rm min}>d-2$), the correction in the lightcone limit is due to conserved currents, in particular, the stress tensor operator. Its OPE coefficient in the OPE $T\subset\OO_\Delta\times \OO_{\Delta}$ of identical scalars is fixed by Ward identities in terms of the scaling dimension $\Delta$ and the central charge $C_T$. This leads to further universal corrections for the OPE data of the double-twist operators in the other channel. 

In two dimensions, the Virasoro vacuum block of a four-point function $\OO_1\times \OO_1\to [T^k]\to \OO_2\times\OO_2$ contains contributions from the stress tensor and an infinite family of composite operators of the schematic form $[T^k]$ for each $k$ which are completely fixed by the symmetries. In higher-dimensional CFTs with a large central charge $C_T$, there are similar composite operators $[T^k]_{\tau,s}$ -- with $\tau$ and $s$ denoting the twist and spin, respectively. A priori, the multi-stress tensor OPE coefficients in the OPE of identical scalar operators, $[T^k]_{\tau,s}\subset \OO_\Delta\times \OO_\Delta$, are not fixed by symmetries in contrast to the two-dimensional case. These operators are, however, ubiquitous in theories with gravity duals since they are related to the exchange of multi-graviton states in the bulk. Therefore in order to understand the emergence of gravity in the bulk from the CFT data on the boundary, these operators play a vital role. It is further interesting to ask if there is a notion of universality in the exchanges of multi-stress tensors in holographic CFTs with large $C_T$ and a large gap in the spectrum of higher-spin single trace operators. 

An important case where the exchange of these multi-stress tensors is expected to dominate compared to that of generic operators is when considering heavy states. This is so because the OPE coefficients of multi-stress tensors $[T^k]$ in a scalar OPE $\OO_\Delta\times\OO_\Delta$ scale like $\Delta^k$ for large $\Delta$. An extreme example of this is when the heavy states have dimension $\Delta$ of order $C_T$. Such heavy states are expected to thermalize in holographic CFTs and according to the AdS/CFT dictionary, thermal states on the boundary are dual to black holes in the bulk. Correlation functions of light operators in heavy states therefore provide a possible window into one of the most interesting questions in the AdS/CFT correspondence, the physics of black holes. 

In two dimensions, the heavy-heavy-light-light (HHLL) Virasoro vacuum block was found in \refs{\FitzpatrickVUA,\FitzpatrickZHA} and contains a wealth of information that can be used to shed light on black hole information loss and the thermalization of heavy states, entanglement entropy and much more \refs{\HartmanEE\FaulknerYIA\AsplundRE\CaputaQE\HijanoRLA\HijanoQJA\FitzpatrickFOA\FitzpatrickDLT\AnousBHC
\FitzpatrickIVE\ChenEE\ChenESC\FitzpatrickMJQ\ChenNVBIP\FaulknerHLL\CotlerZFF-\CollierEXN}. In four dimensions\foot{Similar results also holds for $d>2$ with $d$ even.}, recent progress has been made in studying the contribution of multi-stress tensor operators to HHLL correlators, both using conformal bootstrap techniques and the gravitational dual description \refs{\KulaxiziDXO\FitzpatrickZQZ\KarlssonQFI\LiTPF\KulaxiziTKD\FitzpatrickEFK\KarlssonDBD\LiZBA\KarlssonTXU\KarlssonGHX\LiDQM\ParnachevFNA\FitzpatrickYJB-\ParnachevZBR}. In \KarlssonDBD\ following \KulaxiziTKD, it was argued that the contribution of all minimal-twist operators $[T^k]_{\tau_{\rm min},s}$, with $\tau_{\rm min}=2k$ and spin $s=2k+l$ for $l=0,2,4,\ldots$, in holographic CFTs, takes a specific form which is reminiscent to that obtained from the Virasoro vacuum block. It repackages an infinite number of minimal-twist multi-stress tensor OPE coefficients in the HHLL correlator and it is natural to ask if this is governed by an underlying emergent symmetry in the lightcone limit similar to the Virasoro symmetry. 

In this work we present further progress in this direction by studying the HHLL vacuum blocks of two-dimensional CFTs with $\WW_N$ higher-spin symmetry\foot{We will mainly consider $N=3,4$ but the methods used and the structure remains similar for any $N$.}, see \refs{\CastroIW\GaberdielJCA\deBoerSNA\HegdeDQH-\PerlmutterPKF} for related work. The semi-classical vacuum blocks were found for $N=3$ in \refs{\deBoerSNA,\CastroMZA} and for general $N$ in \HegdeDQH. In this case, the charges of the ``light'' operator are large but much smaller than those of the heavy operator which scale with the central charge $c\gg1$. Expanding the $\WW_N$ vacuum blocks in ${q_H^{(i)}\over c}$, where $q_H^{(i)}$ is the spin$-i$ charge of the heavy operator, we find that the result is again similar to the expansion of the Virasoro vacuum block, with a decomposition in terms of composite operators with the correct weight under the global conformal algebra. In particular, when $q_H^{(3)}\sim c\gg q_H^{(i\neq 3)}$, the dominant contributions\foot{Note that it is only the spin-$3$ charge of the ``heavy'' operators that scales with $c$ and, in particular, their scaling dimension is small compared to $c$. We will still refer to these as heavy. It is possible to extend our results to the case when all the charges of the heavy operators are large but we will not attempt to do so since it is the spin-$3$ sector that resembles the stress tensor sector in four dimensions.} are due to composite quasi-primary operators with the schematic form $[W^k]_{2l}$ made out the spin-3 current $W(z)$. The resulting functions, which are linear combinations of products of hypergeometric functions, are also present in the result for the minimal-twist stress tensor sector of the $d=4$ HHLL correlator. This is one of the main motivations for our work.  

We further explicitly compute the first few terms of the $\WW_N$ HHLL vacuum blocks for $N=3,4$ in the limit $q_H^{(3)}\sim c\gg q_H^{(i\neq 3)}$ using an explicit mode calculation. This limit has the advantage that the charges of the light operators are kept fixed as $c\to\infty$ and sheds further light on how the resulting structure that appears in the four-dimensional stress tensor sector of the HHLL correlator could appear from an underlying symmetry algebra. The results agree with those obtained from the expansion of the semi-classical vacuum blocks which assumed that the charges of the light operators were large. This gives further evidence that those results remain true also for finite charge. The mode calculation presented in this work can in principle also be used to compute ${1\over c}$ corrections to the HHLL vacuum blocks.

Focusing on the logarithm of the $\WW_3$ HHLL vacuum block we further show that it satisfies a non-linear differential equation which, in a certain limit, reduces to a cubic equation for the generating function for the sequence of integers given by ${\rm  A085614}$ in \SloaneISS. The $\WW_3$ HHLL vacuum block can also be obtained from a set of diagrammatic rules similar to the Virasoro vacuum block \FitzpatrickFOA. The story can be generalized in the case of the $\WW_4$ HHLL block both in the limit where the spin-4 charge scales with the central charge and is parametrically larger than all other charges and in the limit where the spin-3 charge scales with the central charge and is parametrically larger than the rest of the charges. We expect a similar story to hold for all $\WW_N$ blocks. From a mathematician's point of view, the $\WW_N$ vacuum blocks provide generating functions for several new sequences which can be understood as different generalizations of the Catalan numbers' sequence.  

Further, we examine the stress tensor sector of the four-dimensional HHLL correlator when the conformal dimension of the light operator vanishes, $\DL\to 0$\foot{Note that this is below the unitarity bound. There are, however, certain observables such as the phase shift \refs{\KulaxiziDXO} that do not depend on $\Delta_L$ that one might be able to extract from the $\Delta_L\to0$ limit.}. A similar picture emerges with the relevant sequence of numbers given by the number of linear extensions of the one-level grid partially ordered set (poset) $G[(1^{k-1}),(0^{k-2}),(0^{k-2})]$.\foot{The Catalan numbers are also the numbers of linear extensions of the one-level grid poset $G[(0^{k-1}),(0^{k-2}),(0^{k-2})]$.} We observe the same structure appearing in $d=6,8$ as well. In this case, the sequences of numbers are related to the linear extensions of the $G[({d-2\over 2})^{k-1},(0)^{k-2},(0^{k-2})]$ posets. In the spirit of the two-dimensional cases examined here, one would hope that knowing the algebraic equation satisfied by the generating function of this sequence, would allow the determination of a differential equation satisfied by the all-orders stress-tensor sector of the HHLL correlator in the lightcone limit for  $\DL\to0$. However, to our knowledge, the generating functions of the number of linear extensions of $G[({d-2\over 2})^{k-1},(0)^{k-2},(0^{k-2})]$ are not known.

\subsec{Summary of results}
\noindent Consider a heavy-heavy-light-light (HHLL) four-point function in a two-dimensional CFT with a large central charge $c$ and a higher-spin $\WW_N$ symmetry $\langle \OH(\infty)\OH(1)\OL(z)\OL(0)\rangle$.
The operators $\OH$ and $\OL$ are $\WW_N$ primaries and carry higher-spin charges $q_H^{(i)}$ and $q^{(i)}$, with $i=2,3,\ldots, N$, respectively. Such a four-point function can be decomposed into blocks which contain contributions from a $\WW_N$ primary $\OO$ and all its $\WW_N$-descendants. We define $\GG_N(z)$ as the holomorphic part of the HHLL correlator restricted to the identity block contribution in the direct channel $\OL\times\OL\to 1_{\WW_N}\to \OH\times \OH$. We specify our discussion to the cases $N=3,4$ although it can be generalized to any $N$. 

{\it\subsubsec { HHLL blocks by mode decomposition}} 
\noindent We start by considering the case $N=3$ where the CFT protagonists are the stress tensor $T(z)$ and a spin-$3$ field $W(z)$. $\GG_3(z)$ contains the exchange of all states schematically denoted by 
\eqn\statesexch{
  |\{a_i,b_j\}\rangle := W_{a_1}W_{a_2}\ldots W_{a_n}L_{b_1}L_{b_2}\ldots L_{b_k}|0\rangle - (\ldots)|0\rangle,
} 
where $L_b$ and $W_a$ are the modes of $T(z)$ and $W(z)$, respectively, and the ellipses ensure that these states are mutually orthogonal. In particular, the subsector consisting of only states with modes $L_b$ acting on the vacuum is that of the Virasoro vacuum block and was studied in detail in \FitzpatrickFOA. We are interested in heavy states with a large spin-$3$ charge $w_H\equiv q^{(3)}_H$ with\foot{It is straightforward to extend our results to the case when all the heavy charges are $\OO(c)$ but we will not attempt to do so. See however Appendix A and B.}
\eqn\deflimits{\eqalign{
  &h_H\ll w_H \sim c\to\infty,\cr
  &h, w\ll c,
}}
where $h_H$ and $h$ are the conformal weights of the heavy and light operator, respectively, and $w$ is the spin-$3$ charge of the light operator. The effect of using \deflimits\ is that the dominant contribution to $\GG_3(z)$ is due to states of the form
\eqn\statesexchlim{
  |\{a_i\}\rangle = W_{a_1}W_{a_2}\ldots W_{a_n}|0\rangle - (\ldots)|0\rangle
}
because each $W$-mode will to leading order contribute a factor of $w_H$ when acting on the heavy operators. Inserting the projection on the single mode states $W_{-m}|0\rangle$ in the correlator one finds the $\OO({w_H\over c})$ term of the vacuum block
\eqn\firstorderes{
 \GG_3(z)\Big|_{{w_H\over c}} = {3w w_H\over c}{f_3(z)\over z^{2h}},
}
where $z^{-2h}$ is the disconnected correlator and $f_a(z)$ is an $SL(2;R)$ conformal block given by 
\eqn\sltwor{
  f_a(z)= z^a {}_2F_1(a,a;2a,z).
}
The result in \firstorderes\ is the conformal block due to the exchange of the quasi-primary $W(z)$ and all its descendants under the global conformal group. 

It is useful to recall the behavior of a $d$-dimensional conformal block, $g_{\tau,s}^{(0,0)}(z,\zbar)$, in the lightcone limit $\zbar\to 0$ 
\eqn\lclimitblock{
   g_{\tau,s}^{(0,0)}(z,\zbar)\sim \zbar^{\tau\over 2}f_{{\tau\over 2}+s}(z).
}
In four dimensions, the stress tensor block with $\tau=s=2$ has the same $z$-dependence as \firstorderes\ (as can be seen from \lclimitblock). 

Going back to $d=2$, we consider the $\OO({w_H^2\over c^2})$ contribution to $\GG_3(z)$. This is due to the (unnormalized) states 
\eqn\doublemodestatessum{
  |Y_{m,n}\rangle = \Big[W_{-n}W_{-m}-{(3n+2m)m(m^2-1)(m^2-4)\over 30 (m+n)((m+n)^2-1)}L_{-m-n}\Big]|0\rangle,
}
where the second term ensures that they are orthogonal to the states $L_{-n-m}|0\rangle$. Projecting onto these states one finds that  
\eqn\secorderWthree{
  \GG_3(z)\Big|_{{w_H^2\over c^2}} = \Big[{1\over 2}\Big({3ww_H\over c}f_3(z)\Big)^2-{9w_H^2 h\over 70c^2}w_3(z)\Big]z^{-2h},
}
where $w_3=-14f_3^2+15 f_2f_4$. The resulting simple-looking expression can be decomposed into global conformal blocks of $[W^2]_{2l}$, with weights $h=6,8,\ldots$, with the use of a product formula for hypergeometric functions
found in \KulaxiziTKD. 

Eq.~\secorderWthree\ shows that the vacuum block contribution to the correlation function at quadratic order in the heavy charge expansion can be written as a sum of products $f_af_b$ such that $a+b=6$, where $h=6$ is the weight of the lightest operator $[W^2]_{0}$. In higher, even spacetime dimension a similar picture emerges. In particular it was shown in \refs{\KulaxiziTKD,\KarlssonDBD} that the minimal-twist double-stress tensor contributions to HHLL correlators in four dimensions can be written as $\GG_{d=4}|_{\Delta_H^2/C_T^2}\propto a_{15}f_1f_5+a_{24}f_2f_4+a_{33}f_3^2$, for some $\Delta_L$ dependent coefficients $a_{ij}$. 

Let us now include a spin-$4$ current $U(z)$. With the four-dimensional results quoted above in mind, we consider the $\WW_4$ HHLL vacuum block in the limit where the spin-3 charge is parametrically larger than the rest (this is done in Appendix B). The states \doublemodestatessum\ have a non-vanishing overlap with the single mode states $U_{-m-n}|0\rangle$ and by removing this overlap, one finds that the correction to the $\OO({w_H^2\over c^2})$ term in \secorderWthree\ is proportional to the spin-$4$ charge $u$ of the light operator. The result takes the form 
\eqn\Gfourquadraticres{
\GG_4(z)\Big|_{w_H^2\over c^2} \propto a_{4,15}f_1f_5+a_{4,24}f_{2}f_4+a_{4,33}f_3^2,
} 
with coefficients $a_{4,ij}$ linear in the charges $(h,u)$ of the light operator and quadratic in $w$ due to the first term in \secorderWthree\foot{Whilst the form of the $\GG_4(z)$ at quadratic order matches that of the four-dimensional result (notice the presence of the $f_1 f_5$-term), there is no choice of the charges of the light operators which would yield an exact match.}. 

The results herein, obtained using explicit mode calculations, are in agreement with those for the $\WW_N$ semi-classical vacuum blocks obtained in \HegdeDQH. While the mode calculation becomes tedious at higher orders in ${w_H\over c}$, the expansion of the semi-classical vacuum block is straightforward. Generally, we find that the expansion of the logarithm of the HHLL vacuum block in powers of ${w_H\over c}$ can be written as a linear combination of products of hypergeometric:
\eqn\whtreeresultSum{
  \log\Big(z^{2h}\GG_N(z)\Big) = \sum_{k=1}^\infty\Big({w_H\over c}\Big)^k\sum_{\{i_p\}} b_{N, i_1\ldots i_k}f_{i_1}(z)\ldots f_{i_{k}}(z),
} 
where we have normalized the expression by the (holomorphic) part of the disconnected correlator $z^{-2h}$. $i_p$ are integers such that $i_1+\ldots+i_k=3k$ and the coefficients $b_{N, i_1,\ldots i_p}$ are linear in the charges $q^{(i)}$ of the light operator\foot{Although the form of the $\WW_N$ vacuum block expansion resembles that of the four-dimensional one, there is no value of $N$ that would yield an exact match.}.

{ \it\subsubsec { Generalized Catalan numbers and differential equations}}

\noindent It is instructive to examine the behavior of the vacuum blocks when $z\to 1$.
Similarly to the case of the Virasoro vacuum block, we observe that the logarithm of the $\WW_N$ vacuum block, with one of the heavy charges  $q_H\sim c\to\infty$ and all other charges fixed and parametrically smaller, has the following behavior in the limit $z\to 1$: 
\eqn\logbeh{
  \log(\GG_N(z))\sim B_N\Big(q^{(i)},{q_H\over c}\Big)\log(1-z),
}
where the function $B_N$ is linear in the light charges $q^{(i)}$ and can be perturbatively expanded in ${q_H\over c}$. This behavior is non-trivial since generally a product of $k$ functions $f_a$ is a $k$-th order polynomial in $\log(1-z)$ with coefficients that are rational functions of $z$. 

For the Virasoro case, the corresponding function $B_2$ is the generating function of the Catalan numbers. For $\WW_3$ in the limit $w_H\sim c\to\infty$, with the other charges parametrically smaller and for certain values of the ratio of the charges of the light operator, we find that $B_3$ satisfies a cubic  equation. Inspired by it, one can construct similarly to the Virasoro case, a cubic differential equation satisfied by $\FF_3\equiv\log \GG_3$ with \deflimits. We present it below in the case $h=3w$:
\eqn\difeqwtSum{
{1\over 6w}{d^{3}\over{dz^3}}\FF_3(z) =-{1\over 54 w^3}\left({d\over{dz}}\FF_3(z)\right)^3 +{1\over {6 w^2}}\left({d^2\over{dz^2}}\FF_3(z)\right) \left({d\over{dz}}\FF_3(z)\right)+{{2x}\over{(1-z)^3}},
}
\noindent where $x=6{w_H\over c}$.  We also derive diagrammatic rules for the $\WW_3$ HHLL vacuum block satisfies.

We also consider the $\WW_4$ HHLL vacuum block in Appendix B. We study its behavior in the region $z\sim 1$ in two different cases; when the spin-4 charge, $u_H\sim c\gg 1$ while $h_H, w_H \ll c$ and when the spin-3 charge scales with c, $w_H\sim c \gg 1$ but  $u_H,h_H\ll c$. In both cases the logarithm of the HHLL vacuum block behaves as $\FF_4\sim \log{(1-z)}$ in the limit $z\to 1$. In the former case, the generating function $B_4$ defined according to \logbeh, satisfies a quartic equation for four different choices of the ratio $h/u$. In particular, when $h=5u$ one can show that $\log \GG(z)$ solves a differential equation whose form is inspired by the algebraic equation satisfied by $B_4$. The situation is similar but slightly more involved when the spin-3 charge, $w_H\sim c$.

Finally, we study the stress tensor sector of the HHLL correlator in $d$-spacetime dimensions in the limit $z\to 1$. In this case, we further have to take the $\Delta_L\to0$ limit in order to remove higher log terms and find that the corresponding sequence of numbers are those of the number of linear extensions of posets $G[({d-2\over 2})^{k-1},(0)^{k-2},(0^{k-2})]$. These are generalizations of the Catalan numbers which can be obtained as the number of linear extensions of the simpler poset $G[(0^{k-1}),(0^{k-2}),(0^{k-2})]$. 

\subsec{Outline}
\noindent Section 2 is devoted to explicit mode calculations of the HHLL vacuum blocks. Specifically, in Section 2.1 we review the Virasoro result and in Section 2.2 we generalize this calculation to the case of the $\WW_3$ HHLL vacuum block. In Section 3, we study the behavior of the HHLL vacuum blocks in the region $z\sim 1$. After a short review of the Virasoro case, in section 3.2 we focus on the $\WW_3$ vacuum block. We observe the appearance of a generalized Catalan sequence, determine its generating function and the algebraic equation the latter satisfies. Inspired by this algebraic equation, we determine a cubic differential equation satisfied by the logarithm of the $\WW_3$ vacuum block for certain ratios of the charges of the light operators. We conclude the discussion of the spin-3 case with new diagrammatic rules for the $\WW_3$ vacuum block expansion. In section 3.4 we investigate in a similar manner the stress tensor sector of the four-dimensional HHLL correlator in holographic CFTs. We conclude with a discussion in Section 4. In Appendix A one finds further details on the explicit mode calculations for the $\WW_3$ HHLL vacuum block. In Appendix B we consider the $\WW_4$ HHLL vacuum block. When $w_H$ is the only large charge, we show using the $\WW_4$-algebra that one gets an extension of the $\WW_3$ result which takes a form similar to that of the stress tensor sector of the HHLL correlator in $d=4$. When $u_H$ is the only large charge, we show that the HHLL vacuum block and a specific choice of the light charges is again governed by a generalization of the Catalan numbers, and that a corresponding non-linear differential equation can be written down analogous to the $\WW_3$ case. A similar albeit more involved story emerges in the $z\to 1$ limit when the only large charge is $w_H$. 

\newsec{HHLL blocks by mode decomposition}
\noindent In this section we perform a mode calculation of $\WW_N$ higher-spin vacuum blocks in two-dimensional CFTs with large central charge. We review the calculation of the Virasoro vacuum block in Section 2.1 following \refs{\FitzpatrickVUA,\FitzpatrickZHA} and extend this to include higher-spin currents in Section 2.2 and Appendix B. The semi-classical vacuum block, for large charges, in $\WW_N$ theories has been calculated in \refs{\deBoerSNA,\CastroMZA} for $N=3$ and in \HegdeDQH\ for general $N$ in the dual bulk theory using a Wilson line prescription. Expanding these known results we find agreement with those obtained from the mode calculation. The calculation of the $\WW_N$ vacuum block using an explicit mode expansion can in principle be extended to include finite central charge as well as finite charges of the external operators. 

\subsec{Review of the Virasoro vacuum block}
\noindent In this section we use the Virasoro modes to explicitly calculate the first terms due to Virasoro descendants of the vacuum following \refs{\FitzpatrickVUA,\FitzpatrickZHA}.

We consider a four point function of pair-wise identical operators $\OH$ and $\OL$ with conformal weight $H$ and $h$, respectively, given by $\langle \OH(\infty)\OH(1)\OL(z)\OL(0)\rangle,$
where we suppress the anti-holomorphic part, have used conformal symmetry to fix the operators at $0,z,1,\infty$ and set $\OH(\infty)= \lim_{z\to \infty}z^{2H}\OH(z)$. The limit that will be considered is $c\to\infty$ with $h$ and ${H\over c}$ fixed. 

 We are interested in the contribution due to Virasoro descendants of the vacuum, i.e.\ states of the schematic form 
\eqn\VirDesc{
    \GG_2(z)=\langle \OH(\infty)\OH(1)\sum_{\{m_i\},\{n_j\}}{L_{-m_1}L_{-m_2}\ldots L_{-m_i}|0\rangle\langle0|L_{n_j}\ldots L_{n_2}L_{n_1}\over \NN_{\{m_i\},\{n_j\}}}\OL(z)\OL(0)\rangle,
}
where $\NN_{\{m_i\},\{n_j\}}$ is a normalization factor and $\GG_2(z)$ is defined as the HHLL correlator restricted to the contribution of the identity block in the direct channel (the subscript $(2)$ here stands for the Virasoro algebra as opposed to $(N)$ for the $\WW_N$). In \FitzpatrickFOA\ an orthogonal basis was constructed in the limit $c\to\infty$ and it was shown how to perform this sum using a recursion relation. The correlator organizes into powers of ${H\over c}$ and we will study the first two terms in this expansion. These are due to single and double mode states respectively.

To begin with, consider the contribution from states of the form $L_{-n}|0\rangle$ in \VirDesc. To calculate this, we need the Virasoro algebra 
\eqn\VirAlg{
  [L_m,L_n] = (m-n)L_{m+n}+{c\over 12}m(m^2-1)\delta_{m+n,0},
}
as well as the action on primary operators 
\eqn\ActLm{
  [L_n,\OO(z)] = z^n[h(n+1)+z\pa]\OO(z). 
}
It is straightforward to evaluate $\langle0|L_n \OO(z)\OO(0)\rangle$ and $\langle \OH(\infty)\OH(1)L_{-n}|0\rangle$ for $n\geq 2$ with the help of \ActLm. We find that
\eqn\actThreeLm{\eqalign{
  \langle0|L_n \OO(z)\OO(0)\rangle &=z^n[h(n+1)+z\pa]z^{-2h} = h(n-1)z^{n-2h}\cr
  \langle \OH(\infty)\OH(1)L_{-n}|0\rangle &= H(n-1).
}
}
The norm of these states is given by the central term
\eqn\normST{
  \NN_{n,n} = \langle L_n L_{-n}\rangle = {c\over 12}n(n^2-1). 
}
Combining the above allows one to obtain the single mode state contribution to the vacuum block 
\eqn\singMode{\eqalign{
  \GG_2(z)|_{{H\over c}} &= z^{-2h}\sum_{n=2}^\infty {12 Hh\over c}{(n-1)\over (n+1)}{z^{n}\over n}= {2Hh\over c}f_2(z)z^{-2h},
} 
}
where the $SL(2,R)$ blocks $f_a$ are given by 
\eqn\defF{
  f_a(z) = z^a {}_2F_1(a,a;2a;z). 
}

Consider now states of the schematic form $L_{-m}L_{-n}|0\rangle$. These are not orthogonal to the single mode states $L_{-m-n}|0\rangle$ since
\eqn\overlapVir{
  \langle L_{m+n}L_{-n}L_{-m}\rangle = (2n-m){c\over 12}m(m^2-1)\neq0. 
}
Removing this overlap one can construct states $|X_{m,n}\rangle$\foot{Note that the states $|X_{m,n}\rangle$ thus defined are not unit normalised.} that are orthogonal to $L_{-m-n}|0\rangle$:
\eqn\doubleMode{
  |X_{m,n}\rangle = \Big[L_{-n}L_{-m}-{\langle L_{m+n}L_{-n}L_{-m}\rangle\over \langle L_{m+n}L_{-m-n}\rangle}L_{-m-n}\Big]|0\rangle,
}
which contribute at $\OO({H^2\over c^2})$ to $\GG(z)$. The contribution of these states can be found from (for details see Appendix A, as well as \FitzpatrickFOA)
\eqn\XCont{\eqalign{
  \langle 0|L_{m}L_n\OL(z)\OL(0)\rangle &= \left[h^2(n-1)(m-1)+hm(m-1)\right]z^{s-2h},\cr
  \langle 0|L_{m+n}\OL(z)\OL(0)\rangle &= h(s-1)z^{s-2h},
}}
where $s=m+n$. With the help of \XCont\ one finds that 
\eqn\xContLight{\eqalign{
  \langle X_{m,n}|\OL(z)\OL(0)|0\rangle &= \Big[h^2(n-1)(m-1)+hm(m-1)\cr
  &-{(2n-m){c\over 12}m(m^2-1)\over {c\over 12}s(s^2-1)}h(s-1)\Big]z^{s-2h}\cr
  &= \Big[h^2(m-1)(n-1)+h{n(n-1)m(m-1)\over s(s+1)}\Big]z^{s-2h}
}
}
as in \FitzpatrickFOA.  Furthermore, keeping only the leading term for large $H$ gives  
\eqn\XContHeavy{
  \langle \OH(\infty)\OH(1)|X_{m,n}\rangle = H^2(n-1)(m-1).
}
The norm of the states $|X_{m,n}\rangle$ in the large-$c$ limit is given by the square of the central terms, {\it{i.e.}},
\eqn\normX{
  \NN_{X_{m,n}} = \langle L_{m}L_n L_{-n}L_{-m}\rangle = \Big({c\over 12}\Big)^2m(m^2-1)n(n^2-1)+\ldots,
}
where the ellipses refer to terms subleading in $c$. Combining the above one finds the contribution of the states $|X_{m,n}\rangle$ to the vacuum block in \VirDesc\ to be 
\eqn\doubleModeVir{\eqalign{
  \GG_2(z)|_{{H^2\over c^2}} &= {z^{-2h}\over 2}\Big({12 Hh\over c}\Big)^2\sum_{m,n=2}^\infty {(m-1)(n-1)\over (m+1)(n+1)}{z^{m+n}\over mn}\cr
  &+ z^{-2h}{72 H^2 h\over c^2}\sum_{m,n=2}^\infty {(m-1)(n-1)\over (m+1)(n+1)}{z^{m+n}\over (m+n)(m+n+1)},
}
}
where we have included a symmetry factor of ${1\over 2}$ due to the exchange symmetry $(m\leftrightarrow n)$. The first line in \doubleModeVir\ comes from the exponentiation of the first term, {\it i.e.}, it is the square of \singMode\ divided by $2$
\eqn\exptermVir{
  \GG_2(z)|_{{H^2h^2\over c^2}} = {1\over 2}\Big({2Hh\over c}f_2\Big)^2z^{-2h}.
}
The second line in \doubleModeVir\ can be written as a sum of products of functions $f_a f_b$ such that $a+b=4$ in the following way 
\eqn\secLineVIr{
  \GG_2(z)|_{{H^2 h\over c^2}} = z^{-2h}{2H^2h\over c^2}\Big[-f_2^2+{6\over5}f_1 f_3\Big]
}
as was noted in \KulaxiziDXO. 

The relative coefficient between the terms in the bracket of \secLineVIr\ is precisely such that in the limit $z\to 1$ the coefficient in front of $\log^2(1-z)$ vanishes and \secLineVIr\ behaves as 
\eqn\secLineLogSq{
  \GG_2(z)|_{{H^2 h\over c^2}} \lzone \log(1-z). 
}
In \FitzpatrickFOA\ it was noticed that this behavior persists to all orders, {\it i.e}, the coefficients of all the $\log^p{(1-z)}$ with $p>1$ vanish in the limit $z\to 1$ and hence $\GG_2(z)$  has a simple logarithmic behavior in this limit. Moreover, the authors of \FitzpatrickFOA\ observed that the coefficients in front of the $\log(1-z)$ terms at each order in ${H \over c}$ form the Catalan numbers' sequence. In the following sections we will see a similar statement being true for $\WW_{N=3,4}$ vacuum blocks\foot{We expect this to be true for arbitrary $N$.}.

\subsec{$\WW_3$ vacuum block}
\noindent In an effort to elucidate the connection between the structure of the vacuum block in the ${H\over c}$ expansion and the underlying symmetry algebra, we consider now a $2$d CFT with a spin-$3$ current $W(z)$. The spin-$3$ modes are defined by 
\eqn\spinthreeMode{
  W(z) = \sum_{n}W_n z^{-n-3},
}
and satisfy the $\WW_3$ algebra 
\eqn\WthreeAlgebra{\eqalign{
  [L_m,W_n] &= (2m-n)W_{m+n},\cr
  [W_m,W_n] &= {c\over 360}m(m^2-1)(m^2-2^2)\delta_{m+n}+\cr
   &+(m-n)\Big[{1\over 15}(m+n+3)(m+n+2)-{1\over 6}(m+2)(n+2)\Big]L_{m+n}\cr
   &+{16\over 22+5c}(m-n)\Lambda_{m+n},
}
}
where $\Lambda_{m}=\sum_p :L_{m-p}L_p:-{3\over 10}(m+2)(m+3)L_m$. The spin-$3$ current $W(z)$ is a primary operator normalised so that $\langle W(z)W(0)\rangle={c\over 3 z^6}$. Note that the non-linear terms in \WthreeAlgebra\ are suppressed in the large-$c$ limit. 

We will study the $\WW_3$ vacuum block $\GG_{3}$ contribution to the four point function of pairwise identical scalars $\OH$ and $\OL$. These are $\WW_3$ primaries and have conformal weights $H$ and $h$, as before, as well spin-$3$ charges $\pm w_H$ and $\pm w$, respectively, with the following scaling as $c\to\infty$: \foot{In \AfkhamiJeddiIDC\ it was shown that unitary representations have weight $\tilde{h}\sim c$ and therefore neither the heavy nor the light operators we consider are unitary.}
\eqn\scaling{w_H\gg H,h,w, \qquad {w_H\over c}=\, {\rm fixed}.
}
As we will see, the contribution from the pure Virasoro modes considered in the previous section is suppressed compared to that containing the spin-3 charge modes of the ``heavy'' operator and is due to states of the schematic form $W_{-m_1}\ldots W_{-m_i}L_{-n_1}\ldots L_{-n_j}|0\rangle$. To evaluate the contribution of such states explicitly, we need to construct an orthogonal basis using the algebra \WthreeAlgebra\ and find the commutator $[W_m,\OO]$. 

Consider first the commutator $[W_m,\OO]$. This is determined by the singular terms in the OPE 
\eqn\OPEWO{\eqalign{
  W(z)\OO(0)|0\rangle &= z^{-3}W_0|h,w\rangle +z^{-2}W_{-1}|h,w\rangle+z^{-1}W_{-2}|h,w\rangle+\OO(z^0)\cr
  &= z^{-3}w\OO|0\rangle + z^{-2}(\OO_{h+1}+{3w\over 2h}\pa \OO)|0\rangle \cr
  &+ z^{-1}(\OO_{h+2}+{2\over h+1}\pa \OO_{h+1}+{3w\over h(2h+1)}\pa^2\OO)|0\rangle+\ldots,
}
}
where $\OO_{h+1}$ and $\OO_{h+2}$ are quasi-primary operators with conformal weight $h+1$ and $h+2$, respectively, and are given by 
\eqn\defQPW{\eqalign{
  \OO_{h+1}(0)|0\rangle &:= \Big[W_{-1}\OO-{3w\over 2h}L_{-1}\OO\Big]|0\rangle,\cr
  \OO_{h+2}(0)|0\rangle &:= \Big[W_{-2}\OO - {2\over h+1}L_{-1}\OO_{h+1}-{3w\over h(2h+1)}L_{-1}^2\OO\Big]|0\rangle.
}
}
Being quasi-primaries, they satisfy $[L_1,\OO_{h+1}(0)]=[L_1,\OO_{h+2}(0)]=0$ which can be verified using the algebra \WthreeAlgebra. The commutator $[W_n,\OO]$ can be found using translation invariance, multiplying with $\int_{\CC(z)}{dw\over 2\pi i}w^{n+2}$ and using the OPE \OPEWO\ :
\eqn\commWO{\eqalign{
  [W_m,\OO(z)] =& {w(m+1)(m+2)\over 2}z^m\OO(z)+(m+2)z^{m+1}(\OO_{h+1}(z)+{3w\over 2h}\pa\OO(z))\cr
  &+z^{m+2}(\OO_{h+2}(z)+{2\over h+1}\pa \OO_{h+1}(z)+{3w\over h(2h+1)}\pa^2\OO(z)). 
}}

Consider now the contribution to $\GG(z)$ from states $W_{-n}|0\rangle$. In order to calculate $\langle W_{n}\OO(z)\OO(0)\rangle$\foot{We denote $\OL\equiv \OO$ to simplify the notation.}, we note that $\langle \OO_{h+1}(z)\OO(0)\rangle=\langle \OO_{h+2}(z)\OO(0)\rangle=0$ since these and $\OO$ are quasi-primaries with different conformal weights. It follows that only $\OO$ and its global descendants in \commWO\ contribute to $\langle W_{m}\OO(z)\OO(0)\rangle$, leading to
\eqn\actWLight{\eqalign{
  \langle W_{n}\OO(z)\OO(0)\rangle &= z^n \Big[{w\over 2}(n+1)(n+2)+{3w\over 2h}(n+2)z\pa_z+{3w\over h(2h+1)}z^2\pa_z^2\Big]z^{-2h}\cr
  &= {w\over 2}(n-1)(n-2)z^{n-2h},
}}
where the operator at $z$ has spin-$3$ charge $w$ and the operator at $0$ has charge $(-w)$. On the other hand, for the heavy part, one finds that 
\eqn\actWHeavy{
  \langle\OH(\infty)\OH(1)W_n\rangle = {w_H\over2}(n-1)(n-2),
}
where the operator at $z=1$ carries spin-$3$ charge $(-w_H)$ and the one at $z \to \infty$, charge $w_H$. Multiplying \actWLight\ with \actWHeavy, dividing with the norm given by the central term in \WthreeAlgebra\ and summing over $n=3,4,\ldots$, one finds the expected result for the $\WW_3$ vacuum block due to the exchange of a spin-$3$ quasi-primary 
\eqn\contWSingleMode{\eqalign{
  \GG_{3}(z)|_{{w_H w\over c}} &= z^{-2h}{90 w_H w\over c}\sum_{n=3}^\infty {(n-1)(n-2)\over (n+1)(n+2)}{z^n\over n}= {3w_H w\over c}f_3(z)z^{-2h}.
}}

Consider now states of the form $W_{-n}W_{-m}|0\rangle$. These are orthogonal to $W_{-n}|0\rangle$ since $W(0)$ does not appear in the OPE $W(z)W(0)$. On the other hand, the stress tensor appears in this OPE and the overlap $\langle L_{m+n}W_{-n}W_{-m}\rangle$ is non-zero. The overlap can be calculated using the fact that $W(z)$ is a primary field. With the help of the first line in \WthreeAlgebra\ one finds 
\eqn\overlapWL{
  \langle L_{m+n}W_{-n}W_{-m}\rangle = {c\over 360}(3n+2m)m(m^2-1)(m^2-4).
}
Removing this overlap leads to states orthogonal to the single-mode ones
\eqn\WWStates{
  |Y_{m,n}\rangle = \Big[W_{-n}W_{-m}-{(3n+2m)m(m^2-1)(m^2-4)\over 30 (m+n)((m+n)^2-1)}L_{-m-n}\Big]|0\rangle,
}
with norm $\NN_{Y_{m,n}}=\langle Y_{m,n}|Y_{m,n}\rangle=({c\over 360})^2m(m^2-1)(m^2-4)n(n^2-1)(n^2-4)$. The overlap with the double-mode states $L_{-m}L_{-n}|0\rangle$ is suppressed in the large-$c$ limit.

The next step is to compute $\langle W_m W_n \OO(z)\OO(0)\rangle$ using the commutator $[W_n,\OO(z)]$ in \commWO. We find that 
\eqn\WWLight{\eqalign{
  \langle W_m W_n \OO(z)\OO(0)\rangle &= z^n\Big[{w\over 2}(n+1)(n+2)+{3w\over 2h}(n+2)z\pa_z+{3w\over h(2h+1)}z^2\pa_z^2\Big]\langle W_m \OO(z)\OO(0)\rangle\cr
  &+z^{n+1}\Big[(n+2)+{2\over h+1}z\pa\Big]\langle W_m \OO_{h+1}(z)\OO(0)\rangle+\cr
  &+z^{n+2}\langle W_m\OO_{h+2}(z)\OO(0)\rangle.
}
}
To evaluate \WWLight\ one may use the commutators $[W_m,\OO_{h+1}(z)]$ and $[W_m,\OO_{h+2}(z)]$ which are found in Appendix A. Alternatively, recall that the three-point functions $\langle W(z)\OO_{h+1}(z)\OO(z)\rangle$, and $\langle W(z)\OO_{h+2}(z)\OO(z)\rangle$, are fixed by conformal symmetry up to the respective OPE coefficients. This gives  
\eqn\honeCont{\eqalign{
  &z^{n+1}\Big[(n+2)+{2\over h+1}z\pa\Big]\int {dz_3\over 2\pi i} z_3^{m+2}\langle W(z_3)\OO_{h+1}(z)\OO(0)\rangle \cr
  &= \lambda_{W\OO_{h+1}\OO}{m(m-1)(m-2)(h(n-2)+2m+n)\over 6(h+1)}z^{m+n-2h},
}}
where $\lambda_{W\OO_{h+1}\OO}$ is the OPE coefficient of $\OO$ in the OPE $W\times\OO_{h+1}$. Likewise, $\langle W_m \OO_{h+2}(z)\OO(0)\rangle$ is given by
\eqn\ohtwo{
  z^{n+2}\langle W_m \OO_{h+2}(z)\OO(0)\rangle = {\lambda_{W\OO_{h+1}\OO}\over 24}(m-2)(m-1)m(m+1)z^{m+n-2h}. 
}

The OPE coefficients are found with the help of the algebra, \WthreeAlgebra,  by taking the limit $z\to 0$
\eqn\OhsWOhone{\eqalign{
  \langle \OO (z_3)W(z)\OO_{h+1}(0)\rangle &\approx z^{-4}\langle \OO(z_3)W_1(W_{-1}-{3w\over 2h}L_{-1})\OO(0)\rangle\cr
  &= z^{-4}z_3^{-2h}\Big[{h(2-c+32h)\over 22+5c}-{9w^2\over 2h}\Big],\cr
}}
and 
\eqn\OhsWOhtwo{\eqalign{
  &\langle \OO(z_3)W(z)\OO_{h+2}(0)\rangle \cr
  &\approx z^{-5}\langle \OO(z_3)W_2(W_{-2}-{2\over h+1}L_{-1}W_{-1}+{3w\over (h+1)(2h+1)}L_{-1}^2)\OO(0)\rangle\cr
  &=z^{-5}z_3^{-2h}\Big[{8h(6+c+8h)\over 22+5c}-{2\over h+1}{4h(2-c+32h)\over 22+5c}+{36w^2\over (h+1)(2h+1)}\Big]\,.
}}
From \OhsWOhone\ and \OhsWOhtwo\ we deduce that for large-$c$ 
\eqn\OPECoeffs{\eqalign{
  \lambda_{W\OO_{h+1}\OO} &= -{h\over 5}-{9w^2\over 2h},\cr
  \lambda_{W\OO_{h+2}\OO} &= {8h\over 5}+{8h\over 5(h+1)}+{36w^2\over (h+1)(2h+1)}.
}}
Using \honeCont\ and \ohtwo\ and the OPE coefficients given in \OPECoeffs\ to evaluate \WWLight, we find that $\langle Y_{m,n}|\OO(z)\OO(0)\rangle$ is given by 
\eqn\YLight{\eqalign{
  \langle Y_{m,n}|\OO(z)\OO(0)\rangle = &\Big[{w^2\over 4}(m-1)(m-2)(n-1)(n-2)\cr
  &-{h\over 30}{m(m-1)(m-2)n(n-1)(n-2)\over (m+n)(m+n+1)}\Big]z^{m+n-2h},
}}
with $|Y_{m,n}\rangle$ defined in \WWStates. 
The heavy part $\langle\OH(\infty)\OH(1)|Y_{m,n}\rangle$ can be calculated in a similar manner,
\eqn\YHeavy{
  \langle\OH(\infty)\OH(1)|Y_{m,n}\rangle = {w_H^2\over 4}(m-1)(m-2)(n-1)(n-2),
}
in the limit $w_H\gg H$. Multiplying \YLight\ and \YHeavy, dividing by the norm $({c\over 360})^2m(m^2-1)(m^2-4)n(n^2-1)(n^2-4)$ and summing over $m,n=3,4,\ldots$ we determine the contribution of the states $|Y_{m,n}\rangle$ to the $\WW_3$ vacuum block to be:
\eqn\YCont{\eqalign{
  \GG_3(z)|_{{w_H^2\over c^2}} &= {z^{-2h}\over 2}\sum_{m,n=3}^\infty \Big[\Big({90w_H w\over c}\Big)^2{(m-1)(m-2)(n-1)(n-2)\over (m+1)(m+2)(n+1)(n+2)}{1\over mn}\cr
  &-{540 w_H^2 h\over c^2}{(m-1)(m-2)(n-1)(n-2)\over (m+1)(m+2)(n+1)(n+2)}{1\over s(s+1)}\Big]z^{s},
}
}
where $s=m+n$.
The first line in \YCont\ is the exponentiated term analogous to the Virasoro case:
\eqn\WWExp{
  \GG_3(z)|_{{w_H^2 w^2\over c^2}} = {1\over 2}\Big({3w_Hw\over c}f_3\Big)^2z^{-2h}, 
}
while the second line can be summed to 
\eqn\WWNew{
  \GG_3(z)|_{{w_H^2 h\over c^2}} = -{9w_H^2 h\over 70c^2}w_3(z)z^{-2h},
}
where $w_3(z)$ is a sum of products $f_af_b$ with $a+b=6$:
\eqn\defW{\eqalign{
  w_3(z)\equiv& -14f_3^2(z)+15f_2(z)f_4(z)\cr
   =&4200\sum_{m,n=3}^\infty{(m-1)(m-2)(n-1)(n-2)\over (m+1)(m+2)(n+1)(n+2)}{z^s\over s(s+1)}.
}}
Similar to the Virasoro case, it is easy to verify that the non-exponentiated term $w_3(z)$ behaves as $\log (1-z)$ when $z\to 1$.

We can also calculate the contribution to the $\WW_3$ vacuum block from states of the form $\Big[L_{-m}W_{-n}-{\langle W_{m+n}L_{-m}W_{-n}\rangle\over \langle W_{m+n}W_{-m-n}\rangle}W_{-m-n}\Big]|0\rangle$. This results in a term that contributes to exponentation and takes the form $\propto {w_H H w h\over c^2}f_2f_3$, as well as a term $\propto {w_H H w\over c^2}(f_1f_4-{7\over 9}f_2f_3)$. Such terms are subleading in the limit $w_H\gg H$  (see Appendix A.1 for further details).

\newsec{Generalized Catalan numbers and differential equations}

\noindent In this section we study the logarithm of the correlator defined by $\FF_N \equiv\log \GG_N$. We start by reviewing the behavior of the logarithm of the Virasoro vacuum block, $\FF_2=\log \GG_2$, in the limit $z\to 1$, the appearance of the Catalan numbers's sequence, and the  differential equation satisfied by $\FF_2$, following \FitzpatrickFOA. Next, we focus on the case $N=3$ where a very similar story emerges. Besides a certain generalization of the Catalan sequence, we also find a set of diagrammatic rules governing the expansion of the $\WW_3$ vacuum block along with a differential equation satisfied by $\FF_3$ for certain ratios of the values of the charges of the light operators. We also consider the logarithm of the stress-tensor sector of the four-dimensional correlator in the lightcone limit, which we denote by $\GG_{d=4}$ and $\FF_{d=4}$ respectively. We investigate the behavior in the limit $z \to 1$ and observe similarities with the two-dimensional cases when $\DL\to 0$.

\subsec{The Virasoro vacuum block}
\noindent In \FitzpatrickFOA\ it was shown how one can derive a differential equation satisfied by the logarithm of the Virasoro vacuum block, by studying its behavior in the $z\to 1$ limit. Expanding $\FF_2$ in powers of $h_{H}/c$ the authors of \FitzpatrickFOA\ observed that $\FF_2$ behaves logarithmically when $z\to 1$. Furthermore they noticed that the sequence of the numerical coefficients multiplying the logarithm at the each order forms the sequence of Catalan numbers given by $c_{2,k}$:
\eqn\catnumb{c_{2,k}={{\Gamma(2k-1)}\over{\Gamma(k)\Gamma(k+1)}},\qquad k\geq 1.
}
\noindent These numbers are generated by the following generating function
\eqn\catalangenfun{B_2(x)=\sum_{k=1}^{\infty}c_{2,k}x^{k}={{1-\sqrt{1-4x}}\over{2}},
}
\noindent which satisfies 
\eqn\virasororec{B_2(x)=B_2(x)^2+x.
} 
\noindent The Catalan numbers $c_{2,k}$ are known to appear in various problems in combinatorics. Here we would like to point out that they can also be understood as the numbers of linear extensions of one-level grid posets\foot{Partially ordered sets (posets) have a notion of ordering between some of the elements but not necessarily all of them. A linear extension of a partial ordering is a linear extension to a totally ordered set where all the elements are ordered in such a way that the original partial ordering is preserved.} $G([0^{k-1}],[0^{k-2}],[0^{k-2}])$, for $k\geq 1$. Generally, one-level grid-like posets $G[{\bf v},{\bf t},{\bf b}]$, where ${\bf v}=(v_1,\ldots, v_n)$, ${\bf t}=(t_1,\ldots, t_{n-1})$ and ${\bf b}=(b_1,\ldots, b_{n-1})$, can be represented with Hasse diagrams of the following type:

\ifig\LABEL{ Posets denoted by $G([0,0,0,0],[0,0,0],[0,0,0])$ and $G([1,0,2],[1,1],[2,1])$, respectively.} {\epsfxsize4.5in\epsfbox{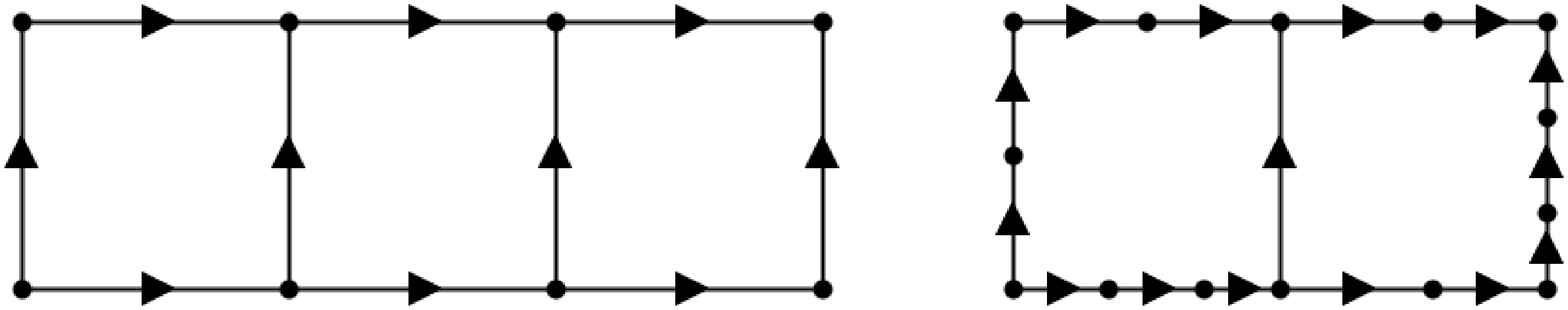}}

\noindent  The numbers $v_i$ denote the number of nodes in the $i$-th vertical edge, $t_i$ denote the number of nodes in the $i$-th top edge and $b_i$ denote the number of nodes in the $i$-th bottom edge, with the endpoints excluded. The Catalan numbers are the numbers of linear extensions of posets of the type depicted in the left Hasse diagram of Fig.~1. 

The logarithm of the correlator $\FF_2(z)=\log\GG_2(z)$ when $z\to 1$ therefore behaves as
\eqn\virbeh{\FF_2(z)\lzone -2h B_2(x) \log(1-z),
} 
\noindent with $x= 6 {h_H\over c}$. Inspired by \virasororec\ and \virbeh\ the authors of \FitzpatrickFOA\ find a differential equation satisfied by $\FF_2(z)$ for all $z$:
\eqn\recdifvir{{1\over {2h}}{d^2\over dz^2}\FF_2(z)={1\over {4h^2}}\left({d\over dz}\FF_2(z)\right)^2 +{x\over{(1-z)^2}}.
}

\subsec{The $\WW_3$ vacuum block}

\noindent Here we uncover a similar story for the $\WW_{3}$ vacuum block $\GG_3$. Expanding in powers of ${w_H\over c}$,
\eqn\fexpwt{\log{\GG_3}\equiv \FF_3(z)=\sum_{k=0}^{\infty}\left({w_H\over c}\right)^{k}\FF_3^{(k)}(z),}
\noindent with
\eqn\fzero{\FF_3^{(0)}(z)=-2h \log(z),
}
and using the exact expression known for the $\WW_3$ vacuum block (see for example eq.~(4.24) in \PerlmutterPKF) one finds that
\eqn\listtwt{\eqalign{\Bigg\{\lim_{z\to 1}&\left(-{\FF_3^{(k)}(z)\over{6^{k+1}\log(1-z)}}\right) \Bigg|k=1,2,\ldots \Bigg\}=\cr
&w\times \big\{1, n, 16, 35 n, 768, 2002 n, 49152, 138567 n ,\ldots\big\},
}}
\noindent where we set $n\equiv h/w$. $\FF_3$ in the limit $z\to 1$ is given by 
\eqn\corder{\FF_3(z)\lzone -6w \log(1-z)B_3(x,n),
}
where $B_3(x,n)$ is the generating function of the sequence \listtwt\ 
\eqn\genfun{B_3(x,n)=\sum_{k=1}^{\infty}c_{3, k}x^{k}={1\over 6} \left(\sqrt{3} \sin \left({1\over 3} \arcsin\left(6 \sqrt{3} x\right)\right)-n\cos \left({1\over 3} \arcsin \left(6 \sqrt{3} x\right)\right)+n\right).
}

Remarkably, there exist exactly three values of $n$ for which $B_3(x,n)$ satisfies a cubic equation; these are $n=\pm 3$ and $n=0$. For these values of the ratios of the light charges, the $\WW_3$ vacuum block simplifies dramatically; it can be expressed in terms of a single function of $z$ raised to a given power\foot{For other values of $n$ the generating function satisfies a sixth order algebraic equation. As a result writing a differential equation becomes cumbersome.}.

For $n = \pm 3$ the sequence of \listtwt\ reduces to
\eqn\listtwthw{\eqalign{
\Bigg\{\lim_{z\to 1}\left[- \left (\pm {1\over 6}\right)^{k+1} \right. &\left.  {\FF_3^{(k)}(z) \over{ \log(1-z)}}\right] \Bigg|k=1,2,\ldots \Bigg\}=\cr
&= w \times \big\{1, \pm3, 16, \pm105, 768, \pm 6006, 49152, \pm 415701,\ldots \big\}.
}}
\noindent 
Each term in this sequence can be derived from the following formula
\eqn\archk{c_{3, k}={{(\pm 2)^{k-1} (3k-3)!!}\over{k! (k-1)!!}},\qquad k\geq 1.
}
Moreover, one can check that function \genfun\ with $n=\pm 3$ satisfies the following relation
\eqn\pert{B_3(x,\pm 3)=-2 B_3(x,\pm 3)^3\pm 3B_3(x,\pm 3)^2 +x. 
}

\noindent with $x=6{w_H\over c}$. Inspired by \pert\ we search for a cubic differential equation satisfied by $\FF_3(z)$. It is easy to see, using the exact expression for the $\WW_3$ block given for example in eq.~(4.24) of \PerlmutterPKF, that $\FF_3(z,n=3)\equiv \hat\FF_3(z)$ satisfies the following differential equation
\eqn\difeqwt{{1\over{6w}}{d^{3}\over{dz^3}}\hat\FF_3(z) =
-{1\over 54 w^3}\left({d\over{dz}}\hat\FF_3(z)\right)^3 +{1\over {6 w^2}}\left({d^2\over{dz^2}}\hat\FF_3(z)\right) \left({d\over{dz}}\hat\FF_3(z)\right)+{{2x}\over{(1-z)^3}}.
}
When ${h\over w}=-3$ a similar equation can be found by taking $w\to -w$ and $1-z\to {1\over 1-z}$. The case $n=0$ is special and is discussed in Appendix B.




{\it\subsubsec { Diagrammatic rules for the $\WW_3$ block}}

\noindent Here we formulate diagrammatic rules for computing the logarithm of $\WW_3$ vacuum block $\FF_3(z)=\log \GG_3(z)$, in the limit where $w_H\sim c\gg 1$ and all other charges are parametrically suppressed. The ratio of the charges of the light operator, $n$, is left arbitrary . The rules are similar to those in \FitzpatrickFOA\ for computing the logarithm of the Virasoro vacuum block. 

We now have cubic and quartic vertices and the exchanged states are modes of the stress tensor and spin-3 current, which we refer to collectively as currents. The only relevant diagrams in the limit we consider, are those where a single propagator connects to the light operator $\OO_L$.  The rules can be stated as follows:

1. Label the $k$ initial currents connected to operator $\OO_H$ with integers $a_1, a_2, \ldots, a_k$.

2. Draw all diagrams where the $k$ initial currents combine via 3-pt and 4-pt vertices to become a single current, which connects with the light operators.

3. For each propagator define its momentum $p$ as the sum of the $a_i$ flowing through it. Momentum is conserved at vertices. Each propagator comes with a factor
$$ {1\over {(p+1)(p+2)}}.$$

4. For each vertex coupling a current of momentum $a_i$ to the external operator $\OO_H$, include a factor of 
$$ {w_H\over \sqrt{c}}(a_i-1)(a_i-2). $$

5. For each vertex coupling a current of momentum $p$ to the external operator $\OO_L$, include a factor of 
$$ {1\over {6\sqrt{c}}} \left((-1)^k (h-3w)+h+3w\right)(p-1)(p-2).$$
\ifig\LABEL{Vertices denoting the coupling of an exchanged current with the external states $\OO_H$ and $\OO_L$, respectively.} {\epsfxsize3.0in\epsfbox{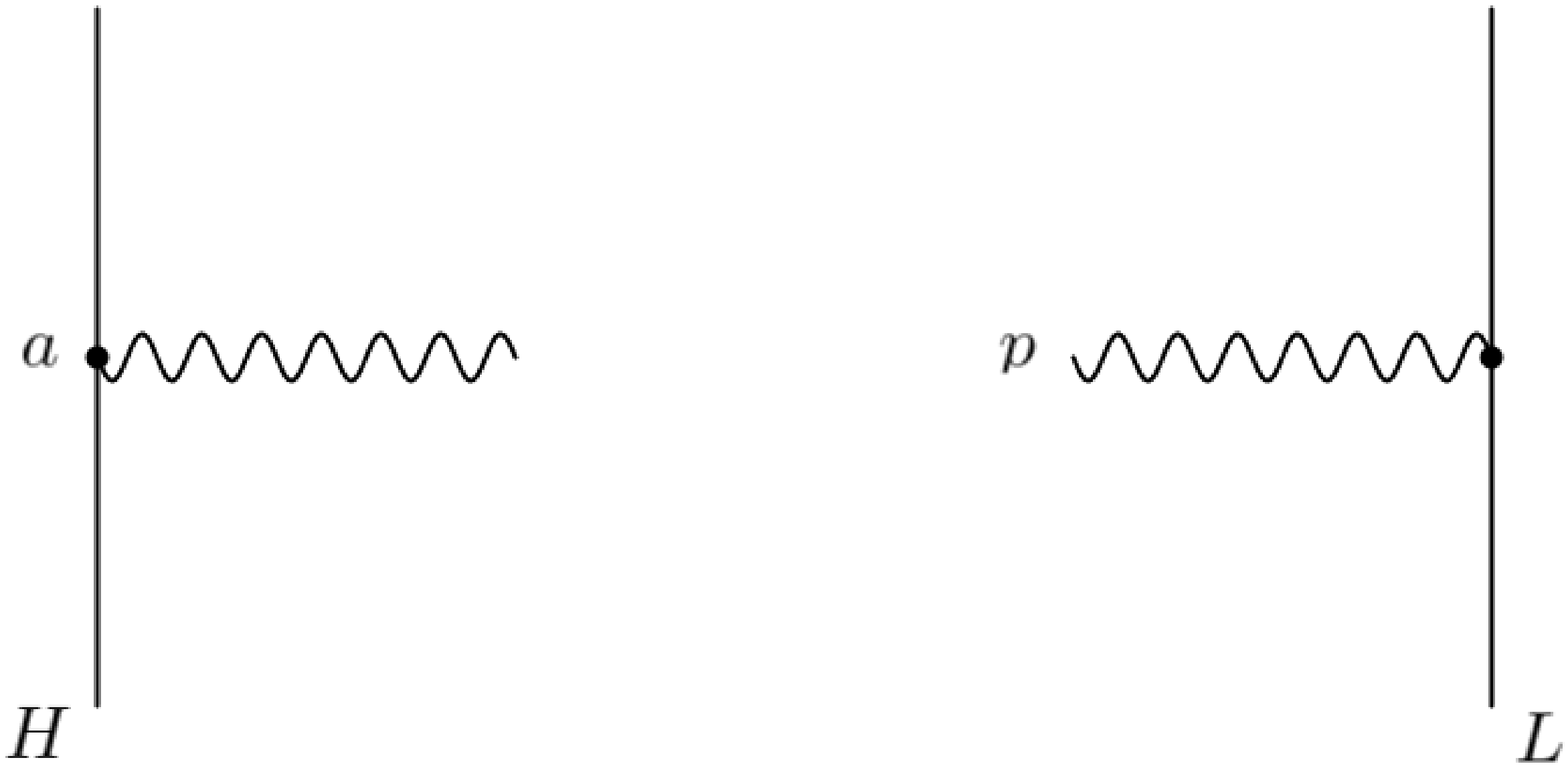}}

6. For each 4-current vertex, include a factor of $-{2/ 3c}$. For each 3-current vertex, where two currents carry momentum $m$ and $n$, while the third current carries momentum $m+n$ (see fig.~3), include a factor of
$${1\over \sqrt{c}}(m+n+2).$$
\ifig\LABELo{Vertices denoting 3-pt and 4-pt coupling of currents, respectively.} {\epsfxsize3.0in\epsfbox{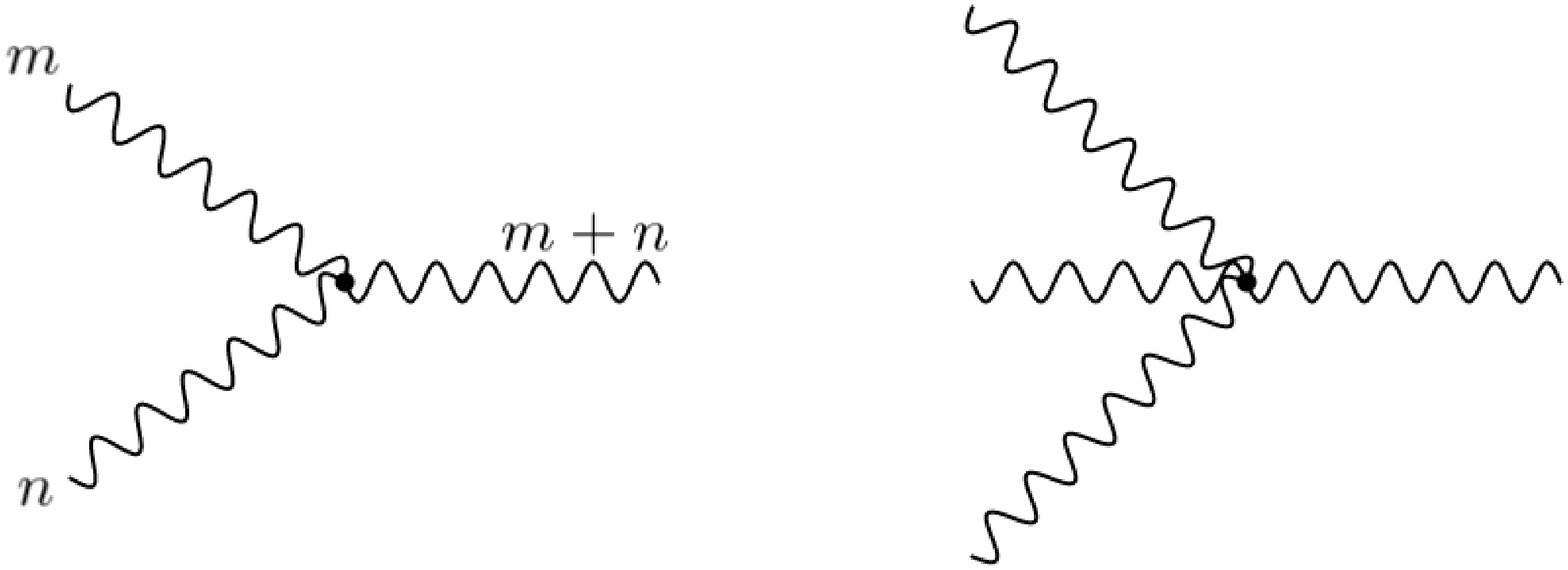}}

7. Take the product of the propagators and vertices and then multiply the result by
$$ {36^{k}\over{k!}}{z^s\over{s(s-1)(s-2)}},$$
where $s=\sum_{i=1}^{k}a_{i}$.

8. Sum the resulting tree diagrams over all $a_i$ from 3 to $\infty$ to obtain the ${w_H^{k}\over c^{k}}$ term in $\FF(z)|_{\WW_3}$.

At orders $w_H/c$ and $w_H^2/c^2$ there is just one diagram to take into account, while at order $w_H^3/c^3$ there are two different types of diagrams. This way, one obtains the expansion of the logarithm of $\WW_3$ vacuum block, which is given by eq.~(4.24) in \PerlmutterPKF .\foot{We explicitly checked this up to $\OO(w_H^4/c^4)$.} 
\ifig\LABEL{Diagrams at orders $w_H/c$ and $w_H^2/c^2$, respectively.} {\epsfxsize3.0in\epsfbox{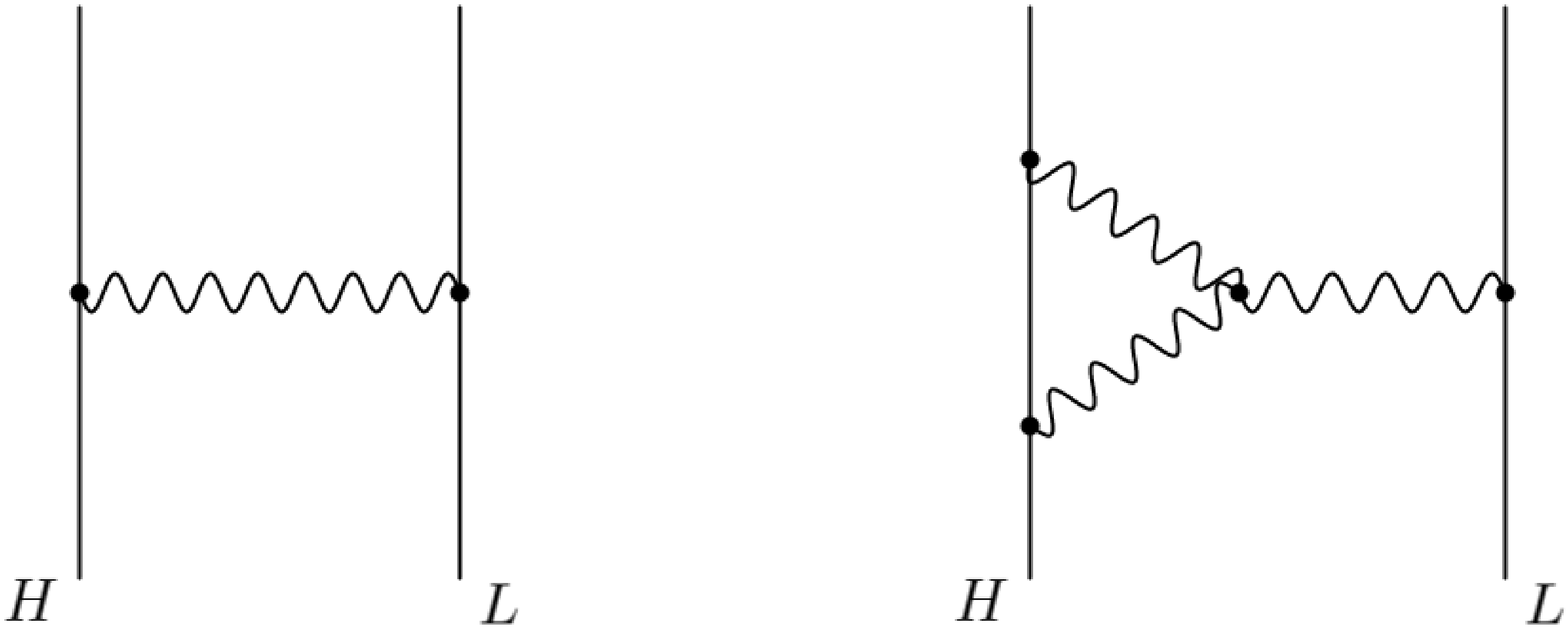}}
\ifig\LABEL{Diagrams at order $w_H^3/c^3$.} {\epsfxsize3.0in\epsfbox{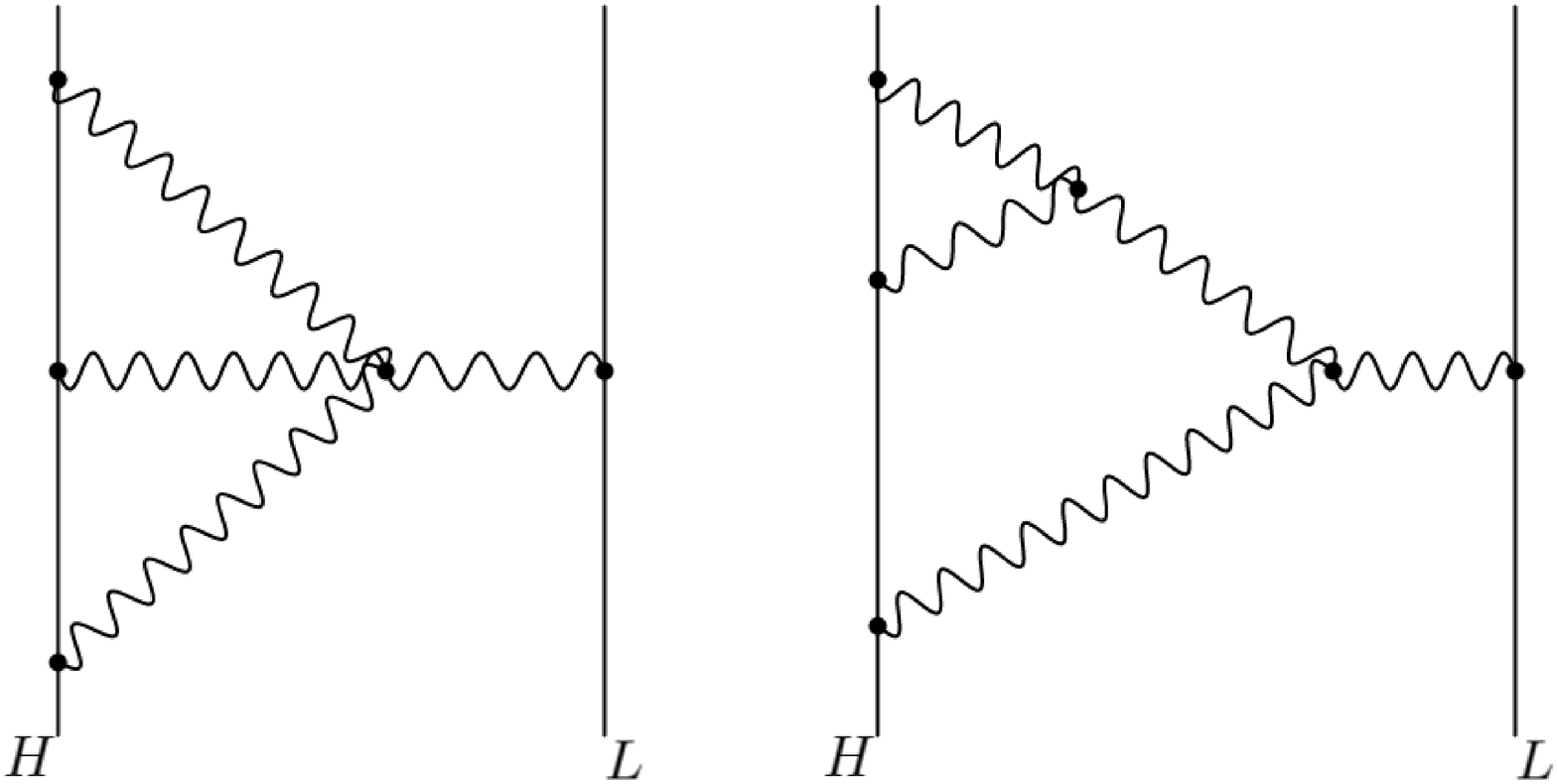}}

\subsec{Stress tensor sector in $d=4$}

\noindent The stress tensor sector of the HHLL correlator in four-dimensional spacetime and in the lightcone limit $(\zbar \to 0)$ is given according to \KarlssonDBD\  by
\eqn\fd{\GG_{d=4}(z,\zbar)={1\over{\left(z \zbar\right)^{\DL}}}\left(1+\sum_{k=1}^{\infty}\mu^{k}\zbar^{k}\GG_{d=4}^{(k)}(z)\right),
} 
\eqn\gk{\GG_{d=4}^{(k)}(z)=\sum_{\{i_{p}\}}a_{i_1 \ldots i_k}f_{i_1}(z)\ldots f_{i_k}(z),
}
\noindent where the sum goes over all sets of $\{i_p\}$ with $i_p \leq i_{p+1}$ and $a_{i_1 \ldots i_k}$ coefficients that depend on $\DL$, and the expansion parameter $\mu$ is given by
\eqn\mudef{\mu\equiv{160\over 3}{\DH\over C_T}.
}
Explicit expressions for $\GG_{d=4}^{(k)}$ with $k=1,2,3$ are given in \KarlssonDBD . There it was also shown that $\GG_{d=4}(z,\zbar)$ can be written as
\eqn\exp{\GG_{d=4}(z,\zbar)={e^{\DL \FF_{d=4}(z,\zbar)},}}

\noindent $\FF_{d=4}(z,\zbar)$ being of $\OO(1)$ in the limit $\DL \to \infty$ and which can be expanded as follows
\eqn\fexp{\FF_{d=4}(z,\zbar)=\FF^{(0)}_{d=4}(z,\zbar)+\sum_{k=1}^{\infty}\mu^{k}\zbar^{k}\FF^{(k)}_{d=4}(z).
}
\noindent with $\FF^{(k)}_{d=4}$ being schematically of the same form as the $\GG_{d=4}^{(k)}$ in \gk. For $k=0,1,2,3$ for instance, we have
\eqn\ffunctions{\eqalign{&\FF^{(0)}_{d=4}(z,\zbar)=-\log(z \zbar),\cr
&\FF^{(1)}_{d=4}(z)={1\over 120}f_{3}(z),\cr
&\FF^{(2)}_{d=4}(z) = {(12-5\DL)f_3(z)^2+{15\over 7}(\DL-8)f_2(z)f_4(z)+{40\over 7}(\DL+1)f_1(z)f_5(z)\over 28800(\DL-2)},\cr
&\FF^{(3)}_{d=4}(z)= b_{117}f_{1}^2(z)f_7(z)+b_{126}f_1(z)f_2(z)f_6(z)+b_{135}f_1(z)f_3(z)f_5(z)\cr
  &+b_{225}f_2^2(z)f_5(z)+b_{234}f_2(z)f_3(z)f_4(z)+b_{333}f_3^3(z),
}}
\noindent where
\eqn\fthreefunction{\eqalign{
  b_{117}&= {{5(\DL +1)(\DL +2)}\over {768768(\DL -2)(\DL -3)}},\cr
  b_{126}&= {{5(5\Delta_L^2 -57\DL -50)}\over{6386688(\DL -2)(\DL -3)}},\cr
  b_{225}&= -{{7\Delta_L^2 - 51\DL -70}\over {2903040(\DL -2)(\DL -3)}},\cr
  b_{135}&= -{11\Delta_L^2-19\DL-18\over 1209600(\DL-2)(\DL-3)},\cr 
  b_{234}&= {(\DL-2)(\DL+2)\over 1209600(\DL-2)(\DL-3)},\cr
  b_{333}&= {7\Delta_L^2-18\DL-24\over 2592000(\DL-2)(\DL-3)}.
}}
\noindent Inspired by the two-dimensional case, we consider the $\FF^{(k)}_{d=4}(z)$ in the limit $z\to 1$. We observe that all terms proportional to $\log^{i}(1-z)$ with $i\geq 2$ vanish in this limit as long as $\DL\to 0$. In this special case, one can show that 
\eqn\listt{\Bigg\{\lim_{z\to 1,\DL \to 0} {{(-4)^{k}(k!)\FF^{(k)}_{d=4}(z)}\over{\log(1-z)}} \Bigg |k=1,2,3,4,5,\ldots \Bigg\} =\big\{1, 1, 6, 71, 1266,\ldots \big\}.
}
\noindent The sequence of numbers in the \listt\ is known as the number of linear extensions of the one-level grid poset $G[(1^{k-1}),(0^{k-2}),(0^{k-2})]$, for $k\geq 1$, given by {\rm A274644} in \SloaneISS . As an example, the $k=5$ case is represented by the Hasse diagram in Fig.~2.
\ifig\LABEL{ The poset denoted by $G([1,1,1,1],[0,0,0],[0,0,0])$.} {\epsfxsize3.0in\epsfbox{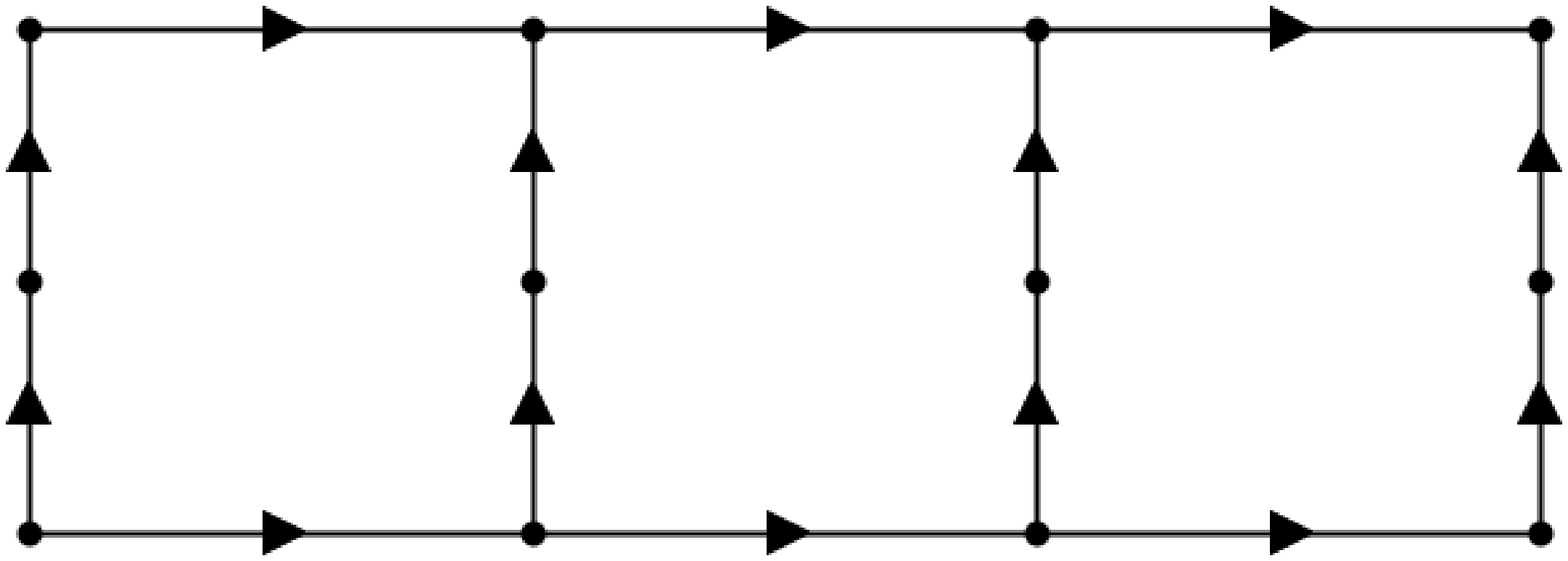}}
\noindent We do not explicitly discuss it here but the relevant posets in even number of dimension $d$ are $G[({d-2\over 2})^{k-1},(0)^{k-2},(0^{k-2})]$. The generating functions and the general formulas for the numbers of linear extensions of posets $G[({d-2\over 2})^{k-1},(0)^{k-2},(0^{k-2})] $ are not (currently) known.

\newsec{Discussion}
\noindent We  consider the  $\WW_N$ vacuum block contributions to heavy-heavy-light-light correlators in two-dimensional CFTs with higher-spin symmetries. 
We perform explicit mode calculations for $\WW_3$ and $\WW_4$ blocks and show that they reproduce the semi-classical vacuum blocks whose explicit form can be 
found in e.g. \HegdeDQH. 
We observe that  terms in the expansion of these blocks in powers of $(q_H^{(i)}/c) $
satisfy the suitably modified ansatz
which was used to compute the stress tensor sector of the $d=4$ HHLL correlator in \KarlssonDBD.

The HHLL Virasoro vacuum block is governed by the Catalan numbers whose generating function satisfies a quadratic  equation allowing the construction of a non-linear differential equation for the logarithm of the vacuum block \FitzpatrickFOA. 
We show that the $\WW_3$ and $\WW_4$ HHLL vacuum blocks are governed by  generalizations of the Catalan numbers; 
for certain values of the  light operator charges, their generating functions satisfy cubic and quartic algebraic equations respectively. 
We further show that these equations uplift to non-linear differential equations satisfied by the logarithm of the blocks. 
What's more, the leading twist stress tensor sector of HHLL correlators in even number of spacetime dimensions $d$  has the same structure
 in the limit $\Delta_L\to 0$. 
 The relevant generalization of the Catalan numbers is now the number of linear extensions of partially ordered sets $G[({d-2\over 2})^{k-1},(0)^{k-2},(0)^{k-2}]$.
 For $d>2$ the generating functions for these sequences are not known. 

The appearance of the generating function $B_N(x)$  comes from the limit $z\to 1$
of the logarithm $\FF_N$ of the block, where $\FF_N\sim B_N(x)\log(1-z)$.
For example, eq. \pert\ defines generalizations of Catalan numbers; this and similar equations were studied in \LISZEWSKA. 
For the $\WW_3$ case, we observe  that for generic light charges $h$ and $w$, the generating function satisfies a polynomial equation of degree $6$, rather than $3$, which however does not take the form studied in \LISZEWSKA. 
The numbers relevant for the $d=4$ result also do not seem to come from equations of this form;
it would be interesting to understand this better. 

Note that in the $d=4$ case, the logarithm of the minimal-twist stress tensor sector of HHLL correlators, $\FF_{d=4}$, is a rational function of $\Delta_L$ which is $\OO(1)$ for large $\Delta_L$. 
An important difference with the $d=2$ $\WW_N$ result
 is that in the limit $z\to 1$, at $k$-th order in the $\mu\simeq {\Delta_H\over C_T}$ expansion,  $\FF_{d=4}^{(k)}\sim g(\Delta_L)\log^{k}(1-z)$ for some function $g(\Delta_L)$. 
 However, in the limit $\Delta_L\to 0$, we do find that $\FF_{d=4}\sim B_{d=4}(\mu)\log(1-z)$ with $B_{d=4}$ being the generating function of the number of linear extensions of the $G[(1^{k-1}),(0^{k-2}),(0^{k-2})]$ posets
 (this is also the number of Young tableaux with restrictions; similar numbers were recently studied in \BanderierRYT)\foot{A similar story holds in $d$ dimensions with the relevant poset now being $G[({d-2\over 2})^{k-1},(0)^{k-2},(0^{k-2})]$. }.
If we knew an algebraic equation satisfied by $B_{d=4}$, we could perhaps construct  a differential equation whose solution would give the full minimal-twist stress tensor sector in $d=4$ large-N CFTs in the limit $\Delta_L\to 0$. 

Heavy-heavy-light-light $\WW_N$ vacuum blocks where the spin-$3$ charge $q^{(3)}_H\sim c$ and $q^{i\neq 3}_H\ll c$ take a form similar to the minimal-twist stress tensor sector in four spacetime dimensions. 
In both cases, at order $({q^{(3)}_H\over c})^k$ in $d=2$ and 
order $\mu^k \simeq ({\Delta_H\over C_T})^k$ in $d=4$, the result is a sum of products $f_{a_1}f_{a_2}\ldots f_{a_k}$ with $a_1+a_2+\ldots +a_k=3k$. In two dimensions, we have shown how at $k=1,2$ and $N=3,4$, this follows from an explicit mode calculation and the knowledge of the higher-spin algebra. It would be interesting to understand if the $d=4$ minimal-twist stress tensor sector  can also be related to
 an emergent symmetry algebra in the lightcone limit. 
 Recently there have been several works devoted to the lightray operators made out of the stress tensor  and to the study of the algebra of such operators \refs{\HuangFOG\BelinLSR\BeskenSNX\HuangHYE-\KorchemskyLRCFT}. 
 It would be interesting to understand if there is a connection to our work. 
 
In \CotlerZFF\ the Chern-Simons description of pure gravity on $AdS_3$ and on Euclidean BTZ  was 
related to the quantization of a certain co-adjoint orbit of the Virasoro group \AlekseevShatashvili.
In this framework the HHLL Virasoro vacuum block, and corrections to it, can be computed.
 It would be interesting to explore a similar framework in the setup of our paper, where higher-spin currents are present.
 
 \bigskip

{\bf Acknowledgments.} We thank C. Banderier, K-W Huang, P. Kraus and S. Mozgovoy for discussions and correspondence. The work of R.K. and A.P. is supported in part by an Irish Research
Council Laureate Award. The work of P.T. is supported in part by an Ussher Fellowship
Award and by the U.S. DOE grant DE-SC00-17660.

\appendix{A}{Some details on the calculation of the $\WW_3$ block}
We now make explicit the contribution of the operator $\OO$ to the commutator $[W_m,\OO_{h+j}(z)]$. To this end, consider the OPE between two quasiprimaries $\phi_i(z_1)\times\phi_j(z_2)|_{\phi^k}$:
\eqn\opeijk{
  \phi_i(z_1)\times\phi_j(z_2)|_{\phi^k} = \lambda_{ijk}\sum_{p=0}^\infty {a_p(h_i,h_j,h_k)\over p!}{\pa_{z_2}^p\phi^k(z_2)\over (z_1-z_2)^{h_i+h_j-h_k-p}},
}
where $a_p(h_i,h_j,h_k)=(h_i-h_j+h_k)_p (2h_k)_p^{-1}$.
Setting $\phi_{i}(z_1)=W(z_1)$, $\phi_j=\OO_{h+j}(z_2)$, $\phi^k=\OO$ and integrating against $\int_{\CC(z_2)} {dz_1\over 2\pi i} z_1^{m+2}W(z_1)\OO_{h+j}(z_2)$ we find that 
\eqn\commOhres{
  [W_m,\OO_{h+j}(z_2)]|_{\OO} = \lambda_{W\OO_{h+j}\OO}\int_{\CC(z_2)} {dz_1\over 2\pi i} z_1^{m+2}\sum_{n=0}^{j+2} {{a_p(3,h+j,h) \pa_{z_2}^p\OO(z_2)}\over {(z_1-z_2)^{3+j-p} p!}},
}
and performing the integral we find that 
\eqn\commOhresfin{
  [W_m,\OO_{h+j}(z_2)]|_{\OO} = \lambda_{W\OO_{h+j}\OO}\sum_{p=0}^{j+2}{{a_p(3,h+j,h) (m+2)!}\over {  (m+p-j)!(j+2-p)! p!}}z_2^{m+n+p-j}\pa_{z_2}^p\OO(z_2).
}

\subsec{Mixed states $W_{-n}L_{-m}|0\rangle$}
\noindent We now consider the following states 
\eqn\WnLm{
  |A_{m,n}\rangle = L_{-m}W_{-n}|0\rangle - {\langle W_{m+n}L_{-m}W_{-n}\rangle \over \langle W_{n+m}W_{-n-m}\rangle}W_{-m-n}|0\rangle,
}
where (for $c\to\infty$)
\eqn\calsone{\eqalign{
  \langle W_{m+n}L_{-m}W_{-n}\rangle &= (3m+n){c\over 360}n(n^2-1)(n^2-4),\cr
  \langle W_{n+m}W_{-n-m}\rangle &= {c\over 360}(m+n)((m+n)^2-1)((m+n)^2-4),\cr
  \langle W_{n}L_mL_{-m}W_{-n}\rangle &= {c^2\over 12\times 360}n(n^2-1)(n^2-4)m(m^2-1).
}}
Now, one finds that $\langle A_{m,n}|\OL(z)\OL(0)\rangle$
\eqn\AOLOL{\eqalign{
  \langle A_{m,n}|\OL(z)\OL(0)\rangle &= \DD_{L,m}\DD_{W,n}\langle\OL(z)\OL(0)\rangle\cr
  &-{\langle W_{m+n}L_{-m}W_{-n}\rangle \over \langle W_{n+m}W_{-n-m}\rangle}\DD_{W,m+n}\langle\OL(z)\OL(0)\rangle\cr
  &= {1\over 2}(m-1)(n-1)(n-2)whz^{m+n-2h}+\cr
  &+{(m-1)m(n-2)(n-1)n(4+m+3n)\over 2(m+n)(m+n+1)(m+n+2)}wz^{m+n-2h},
}
}
where
\eqn\commQ{\eqalign{
  [Q^{(N)}_m,\OO_{h,q^{(N)}}(z)]|_{\OO_{h,q^{(N)}}} &= q^{(N)}\int_{\CC(z)}{dz_1\over 2\pi i}z_1^{m+N-1}\sum_{p=0}^{N-1}{a_p(N,h,h)\over {(z_1-z)^{N-p} p!}}\pa_z^p\OO_{h,q^{(N)}}(z)\cr
  &=q^{(N)}\sum_{p=0}^{N-1}{a_p(N,h,h)\over p!}{(m+N-1)!\over (N-p-1)!(m+p)!}z^{m+p}\pa_z^p\OO_{h,q^{(N)}}(z)\cr
  &:= \DD_{N,m}\OO_{h,q^{(N)}}(z).
}}

For the heavy part, we keep only the quadratic part in the charges such that 
\eqn\OHOHA{
  \lim_{z_4\to\infty} z_4^{2H}\langle \OH(z_4)\OH(1)|A_{m,n}\rangle = {1\over 2}(m-1)(n-2)(n-1)w_H H.
}

Multiplying \AOLOL\ with \OHOHA\ and dividing by the norm $\langle W_{n}L_mL_{-m}W_{-n}\rangle$ in \calsone, we find that 

\eqn\AcontExp{\eqalign{
  &\sum_{m,n=2}^\infty\lim_{z_4\to\infty} z_4^{2h_H}{\langle \OH(z_4)\OH(1)|A_{m,n}\rangle \langle A_{m,n}|\OL(z)\OL(0)\rangle\over \langle W_{n}L_mL_{-m}W_{-n}\rangle}\Big|_{w_H H w h\over c^2} \cr
  &= {1080w_H H w h\over c^2}z^{-2h}\sum_{m,n=2}^\infty {(m-1)(n-1)(n-2)\over (m+1)(n+1)(n+2)}{z^{m+n}\over mn}\cr
  &= {6w_H H w h\over c^2} f_2f_3,
}
}
which as expected is the ``exponentiated term''. On the other hand, consider 
\eqn\Acont{\eqalign{
  &\sum_{m,n=2}^\infty\lim_{z_4\to\infty} z_4^{2H}{\langle \OH(z_4)\OH(1)|A_{m,n}\rangle \langle A_{m,n}|\OL(z)\OL(0)\rangle\over \langle W_{n}L_mL_{-m}W_{-n}\rangle}\Big|_{w_H H w\over c^2} \cr
  &= {1080w_H H w\over c^2}z^{-2h}\sum_{m,n=2}^\infty {(m-1)(n-1)(n-2)\over (m+1)(n+1)(n+2)}{(4+m+3n)z^{m+n}\over (m+n)(m+n+1)(m+n+2)}\cr
  &\propto {w_H H w \over c^2}(f_1f_4-{7\over 9}f_2f_3).
}
}
Note that in both sums we have trivially extended the summation from $m\geqslant 3$ to $m\geqslant 2$. 

On the other hand, by expanding the vacuum block we find precisely the same structure 
\eqn\vacumWL{\eqalign{
  \langle \OO_{H}(\infty)\OO_{H}(1)\OL(z)\OL(0)\rangle|_{1_{\WW_3}, {w_H H w h\over c^2}} &\propto f_2f_3,\cr
  \langle \OO_{H}(\infty)\OO_{H}(1)\OL(z)\OL(0)\rangle|_{1_{\WW_3}, {w_H H w\over c^2}} &\propto (f_1f_4-{7\over 9}f_2 f_3).\cr
}}

\appendix{B}{$\WW_4$ vacuum block}
\noindent In this appendix we further include a spin-$4$ current and consider the $\WW_4$ algebra. We will show that including a spin-4 current modifies the term proportional to ${w_H^2\over c^2}$ discussed in Section 2. The result can again be written as a sums of the following combination $f_a(z)f_b(z)$, with $a+b=6$. Compared to the case of $\WW_3$, the term proportional to ${w_H^2\over c^2}$ in the vacuum block will now depend also on the spin-4 charge $u$ of the light operator. 

 We denote the spin-$4$ current by $U(z)$ and the external operators carry eigenvalues $\pm u_H$ and $\pm u$. The heavy operator again has a spin-$3$ charge of $\OO(c)$ while the conformal weight $H$ and the spin-$4$ charge are small compared to $w_H$, i.e.\ $H,u_H\ll w_H$. In this limit, there are no new contributions due to the states $U_{-m}|0\rangle$ since they will be proportional to ${u_H u\over c} f_4z^{-2h}$, which is suppressed as $c\to\infty$. The first contribution will appear at $\OO({w_H^2\over c^2})$ and is due to the fact that the modes $|Y_{m,n}\rangle$ are not orthogonal to $U_{-m-n}|0\rangle$. In this section we will therefore study the contribution due to the following states:
\eqn\YTilde{
  |\tilde{Y}_{m,n}\rangle = \Big[W_{-n}W_{-m}-{\langle L_{m+n}W_{-n}W_{-m}\rangle \over \langle L_{m+n}L_{-m-n}\rangle}L_{-m-n}-{\langle U_{m+n}W_{-n}W_{-m}\rangle \over \langle U_{m+n}U_{-m-n}\rangle}U_{-m-n}\Big]|0\rangle.
}
There are two new contributions to $\langle\tilde{Y}_{m,n}|\OO(z)\OO(0)\rangle$ compared to $\langle Y_{m,n}|\OO(z)\OO(0)\rangle$, one is simply that we need to include the last term in \YTilde. The second is a correction to the OPE coefficients $\lambda_{W\OO_{h+1}\OO}$ and $\lambda_{W\OO_{h+2}\OO}$, these pick up a contribution that depends on the spin-$4$ charge $u$ due to the fact that $[W_m,W_{-m}]$ contain the spin-$4$ zero mode $U_0$. Note that the heavy part remains unchanged since $w_H\gg u_H$ and is therefore given by \YHeavy:
\eqn\YTildeHeavy{
  \langle\OH(\infty)\OH(1)|\tilde{Y}_{m,n}\rangle = {w_H^2\over 4}(m-1)(m-2)(n-1)(n-2),
}
and the norm of $|\tilde{Y}_{m,n}\rangle$ is also the same as that of $|Y_{m,n}\rangle$ (to leading order in $c$):
\eqn\normYTilde{
  \NN_{\tilde{Y}_{m,n}}=\langle \tilde{Y}_{m,n}|\tilde{Y}_{m,n}\rangle=({c\over 360})^2m(m^2-1)(m^2-4)n(n^2-1)(n^2-4).
}
We therefore only need to calculate $\langle\tilde{Y}_{m,n}|\OO(z)\OO(0)\rangle$.

The modes $U_m$ of $U(z)$ are defined by 
\eqn\modesU{
  U(z) = \sum_m U_m z^{-m-4},
}
and since $U$ is primary we know that
\eqn\UlComm{
  [L_m,U_n] = (3m-n)U_{m+n}. 
}
Consider now various OPEs of the spin-$3$ and spin-$4$ field \foot{See e.g.\ App A.2 in \RasmussenEUS\ for the $\WW_4$ algebra.}, in terms of quasi-primaries
\eqn\UOPE{\eqalign{
  W(z)W(0) &= {c\over 3 z^6}+{2T(0)\over z^4}+ { \lambda_{WWU} U(0)\over z^2}+\ldots,\cr
  W(z)U(0) &= {\lambda_{WUW}W(0)\over z^4}+\ldots,\cr
  U(z)U(0) &= {c\over 4 z^8} + {2T(0)\over z^6}+\lambda_{UUU}{U(0)\over z^4}+\ldots,
}
}
where $\lambda_{WUW}={3\over 4}\lambda_{WWU}={3\over 4}{4\over\sqrt{3}}\sqrt{(2+c)(114+7c)\over (7+c)(22+5c)}\approx \sqrt{21\over 5}$ and the ellipses denote non-linear terms that will be suppressed when $c\to\infty$. 
From \UOPE, we can derive the commutator of the various modes. Especially, we want to consider $[W_n,U_m]$, $[W_n,W_m]$ and $[U_n,U_m]$. The last one is given by
\eqn\centSpinFour{\eqalign{
  [U_m,U_{n}] &= {c\over 20160}m(m^2-1)(m^2-4)(m^2-9)\delta_{m+n}\cr
  &+{(m-n)\over 1680}\Big[3(m^4+n^4)+4m^2n^2-(2mn+39)(m^2+n^2)\cr
  &+20mn+108\Big]L_{m+n}+\ldots,
}
}
while $[W_n,W_m]|_U$ is given by
\eqn\WWU{
  [W_m,W_n]|_{U} = \lambda_{WWU}{m-n\over 2}U_{n+m},
}
as well as
\eqn\WUW{
  [W_m,U_n]|_{W}={\lambda_{WUW}\over 84}\Big[5m^3+9n-5m^2n-n^3-17m+3mn^2)\Big]W_{m+n}.
}

Using \WWU\ and \WUW, we find that 
\eqn\overlapU{\eqalign{
  \langle U_{m+n}W_{-n}W_{-m}\rangle &={\lambda_{WUW} cm(m^2-1)(m^2-4)\over 30240}\cr
  &\times\Big[-9m+m^3-26n+6m^2n+14mn^2+14 n^3\Big]
}}
and 
\eqn\overlapUtwo{
  \langle U_{m+n}U_{-m-n}\rangle = {c\over 20160}s(s^2-1)(s^2-4)(s^2-9).
}
From the three-point function $\langle U(z_3)\OO(z)\OO(0)\rangle$ and $\lambda_{U\OO\OO}=u$ one finds that
\eqn\actU{
  \langle U_{m+n}\OO(z)\OO(0)\rangle = {u\over 6}(m+n-1)(m+n-2)(m+n-3)z^{m+n-2h}. 
}
Lastly, we need to compute the corrections to the OPE coefficients $\lambda_{W\OO_{h+1}\OO}$ and $\lambda_{W\OO_{h+2}\OO}$. This is similar to the calculation in the $\WW_3$ case and one finds that ($c\to\infty$, $z\to 0$)
\eqn\honewfour{\eqalign{
  \langle \OO(z_3)W(z)\OO_{h+1}(0)\rangle &\approx z^{-4}\langle \OO(z_3)W_1(W_{-1}-{3w\over 2h}L_{-1})\OO(0)\rangle\cr
  &= z^{-4}z_3^{-2h}\Big[-{h\over 5}+\lambda_{WWU}u-{9w^2\over 2h}\Big],
}}
where we used $[W_1,W_{-1}]=\ldots +\lambda_{WWU}U_0$ and that $U_0\OO(0)|0\rangle=u \OO|0\rangle$. Likewise, one finds that 
\eqn\honewtwo{\eqalign{
  &\langle \OO(z_3)W(z)\OO_{h+2}(0)\rangle \approx \cr
  &z^{-5}\langle \OO_{h}(z_3)W_2(W_{-2}-{2\over h+1}L_{-1}W_{-1}+\Big[{3w\over h(h+1)}-{3w\over h(2h+1)}\Big]L_{-1}^2)\OO(0)\rangle\cr
  &=z^{-5}z_3^{-2h}\Big[{8h\over 5}+{8h\over {5(h+1)}}+{{36 w^2}\over{(2h+1)(h+1)}}+2\lambda_{WWU}u-{{8u}\over h+1}\lambda_{WWU}\Big],
}}
to leading order when $c\to\infty$ and using $[W_2,W_{-2}]|_U=\ldots +2U_0$. Putting this altogether gives 
\eqn\YtildeCont{\eqalign{
  \langle \tilde{Y}_{m,n}|\OO(z)\OO(0)\rangle &= \langle Y_{m,n}|\OO(z)\OO(0)\rangle\cr
  &+{u\lambda_{WWU}(m-2)(m-1)m(n-2)(n-1)n z^{m+n-2h}\over 12s(s+1)(s+2)(s+3)}\cr
  &\times(17+2m^2+15n+2n^2+15m+9mn)+\ldots.
}
} 

Given \YtildeCont, \YTildeHeavy\ and \normYTilde, we find the contribution to the vacuum block from the states $|\tilde{Y}_{m,n}\rangle$ proportional to $u$ is given by
\eqn\GammaWFourCont{\eqalign{
  \GG(z)|_{{w_H^2 u\over c^2}}&= {37800w_H^2 u\lambda_{WWU}z^{-2h}\over c^2}\Big[25\tilde{w}_{4}(z)+ 3 w_{3}(z)\Big],
}
}
where $w_3$ is given by \defW\ and $\tilde{w}_4$ is a sum of products of functions $f_af_b$ with $a+b=6$ given by
\eqn\defTildewFour{\eqalign{
  &\tilde{w}_4=3(-f_2f_4+{4\over 3}f_1f_5)=\cr
  &\sum_{m=3}^\infty\sum_{n=3}^\infty 1260 {(m-2)(n-2)(n-1)n(m^2+6(n+2)(n+3)+m(9+4n))\over m(n+2)(n+3)(n+4)s(s+1)(s+2)(s+3)}z^{m+n}.
}}

\subsec{Differential equation for the $\WW_4$ vacuum block}

\noindent Here we study the $\WW_4$ vacuum block, or rather its logarithm, as $z\to 1$. The $\WW_4$ HHLL vacuum block is known exactly. One can find it for instance in eq.~(C.1) of \PerlmutterPKF.  In this case, we can choose to scale the spin-3 charge $w_H$ with the central charge $c$ -- as in Appendix A -- with the hope of uncovering relations similar to those valid for the stress-tensor sector of the four-dimensional correlator in the light cone limit. However, we may also choose to consider the limit $u_H \sim c \gg 1$, with all other charges parametrically smaller. 

Remarkably, $\FF_4(z)$ behaves logarithmically in the limit $z\to 1$ in both cases. A sequence of numbers, the numerical coefficients of $\log{(1-z)}$
in the expansion of the relevant heavy charge can be determined, and a quartic differential equation satisfied by the logarithm of the block for certain ratios of the light charges can be found.

Let us first consider the scaling $u_H\sim c\gg 1$ and expand $\FF_4(z)=\log{\GG_4(z)}$ in powers of $u_H/c$ as $\FF_4(z)=\sum_{k=0}^{\infty}\left({u_H\over c}\right)^{k}\FF_4^{(k)}(z)$ to obtain in the limit $z\to 1$:

\eqn\listtwf{\eqalign{&\Bigg\{\lim_{z\to 1}\left(-{\FF^{(k)}(z)\over{20\times6^{k}\log(1-z)}}\right) \Bigg|k=1,2,\ldots \Bigg\}=\cr 
=u \times \big\{&1,n-7,458-14 n,1001 n-13307,732374-34034 n,1939938 n-31667622, \ldots \big\},
}}
\noindent where we set
\eqn\ration{n={18\over 5}{h\over u}.
}
If $B_4(x,n)$ with $x\equiv 6 {u_H\over c}$ is the generating function of  \listtwf, then $\FF(z)$ behaves in the limit $z\to 1$ as
\eqn\limwf{\FF_4(z)\lzone -20 u \log(1-z) B_4(x,n)
}
There exist four different values of $n$ for which the generating function $B_4(x,n)$ satisfies a quartic equation. These are: $n=\{18,3,-2,-12\}$. 

When $n=18$, we find the following quartic order equation for the generating function:
\eqn\polynomone{B_4(x,18)=36 B_4(x,18)^4-36 B_4(x,18)^3+11 B_4(x,18)^2+x.
}
Inspired by this relation one finds that $\FF_4(z,n=18)\equiv\widetilde\FF_4(z)$ satisfies the following differential equation
\eqn\difrecwfo{\eqalign{\widetilde{\FF}''''(z)=120u\Bigg({{x}\over{(1-z)^4}}+{{9 \widetilde{\FF}'(z)^4}\over{40000 u^4}}-{{9 \widetilde{\FF}'(z)^2 \widetilde{\FF}''(z)}\over{2000 u^3}}+{{3 \widetilde{\FF}''(z)^2+4 {\widetilde\FF}'''(z) {\widetilde\FF}'(z)}\over{400 u^2}}\Bigg),
}}
which reduces to the equation \polynomone\ in the limit $z\to 1$ using \limwf.


When $n=-12$ the generating function $B_4(x,-12)$ satisfies
\eqn\recrelb{B_4(x,-12)=-144 B_4(x,-12)^4-96 B_4(x,-12)^3-19 B_4(x,-12)^2+x,
}
whilst $\FF_4(z,n=-12)\equiv\hat{\FF}(z)$ is a solution of the following differential equation 
\eqn\difrecwfoo{\eqalign{\hat\FF''''(z)=120u\Bigg({{x}\over{(1-z)^4}}-{{9 \hat\FF'(z)^4}\over{10000 u^4}}-{{3 \hat\FF'(z)^2 \hat\FF''(z)}\over{250 u^3}} -{{7 \hat\FF''(z)^2+6 \hat\FF'''(z) \hat\FF'(z)}\over{400 u^2}}\Bigg).
}}

For $n=-2,3$ we find the following quartic order equations for the generating function:
\eqn\wfappdod{\eqalign{n&=3,\qquad B_4(x,3)=-2304 B_4(x,3)^4+384 B_4(x,3)^3-4 B_4(x,3)^2+x,\cr
n&=-2,\qquad B_4(x,-2)=2916 B_4(x,-2)^4+324 B_4(x,-2)^3-9 B_4(x,-2)^2 + x.
}}
In these cases however, the differential equations similarly constructed do not correctly reproduce the vacuum block beyond $z\to 1$ limit. This is analogous to what happens in the case of the $\WW_3$ vacuum block for $h=0$, where the generating function satisfies
\eqn\wtappdod{n=0,\qquad B_3(x,0)=16 B_3(x,0)^3 + x.
}
It is curious that these special cases correspond to values for the ratios of the light charges for which $h< w,u$. 

Let us now consider the case with $w_H\sim c\gg 1$ and the other charges parametrically smaller. For notational simplicity, we will use here the same symbol $\FF_4(z)$. We hope that this will not create any confusion. In this case, $\FF_4(z)$ is expanded as
\eqn\fexpwf{\FF_4(z)=\sum_{k=0}^{\infty}\left({w_H\over c}\right)^{k}\FF^{(k)}(z),
}
with
\eqn\fzerowf{\FF_4^{(0)} = -2h \log(z).
}
Using the exact expression for the $\WW_4$ block one finds that
\eqn\listtwt{\eqalign{\Bigg\{\lim_{z\to 1}&\left({ (-1)^{k+1} \FF_4^{(k)}(z)\over{2^{k+1}3^{2k}\log(1-z)}}\right) \Big|k=1,2,\ldots\Bigg\}=\cr
=w\times \big\{& 1,{{2}\over{45}} (18 n+85 m),10,{{2}\over{81}} (882 n+2785 m),318,\cr
&{{44}\over{3645}} (67158 n+225635 m),13620,\ldots \big\},
}}
\noindent where $n,m$ denote the ratios of the light charges $n={h\over w}$ and $m={u\over w}$, respectively. Notice that in this case ratios of both charges appear as opposed to the previous scaling for which additional simplifications occurred that eliminated $w$. This may be related to the fact that a spin-3 current, having odd spin, does not appear in the OPE of two spin-4 currents.


\listrefs

\bye